\documentclass[review]{elsarticle}

\pdfoutput=1

\usepackage{epstopdf}
\usepackage{fullpage}
\usepackage{amssymb}
\usepackage{amsmath}
\usepackage{color}
\usepackage{graphicx}
\usepackage{caption}
\usepackage{subcaption}
\usepackage{amsmath}
\usepackage{algpseudocode}
\usepackage{algorithm}
\usepackage[titletoc,title]{appendix}

\usepackage{booktabs}
\usepackage[para,online,flushleft]{threeparttable}
\usepackage{color}
\usepackage[table,xcdraw]{xcolor}
\definecolor{light-gray}{gray}{0.85}
\journal{ }

\biboptions{authoryear}

\begin{document}
	
\begin{frontmatter}

\title{A Neuro-Fuzzy Model of Time-Varying Decision Boundaries}

\author{Arash Khodadadi\fnref{mycorrespondingauthor}}
\author{Pegah Fakhari}
\author{Jerome R. Busemeyer}
\address{Indiana University, Department of Psychological and Brain Sciences, Bloomington, IN, United States}
\fntext[mycorrespondingauthor]{Corresponding author. Indiana University, Department of Psychological and Brain Sciences, 1101 E. 10th street, 47405-7007, Bloomington, IN, United States.}

\begin{abstract}
In a recent study, we reported the results of a new decision making paradigm in which the participants were asked to balance between their speed and accuracy to maximize the total reward they achieve during the experiment. The results of computational modeling provided strong evidence suggesting that the participants used time-varying decision boundaries. Previous theoretical studies of the optimal speed-accuracy trade-off suggested that the participants may learn to use these time-varying boundaries to maximize their average reward rate. The results in our experiment, however, showed that the participants used such boundaries even at the beginning of the experiment and without any prior experience in the task. In this paper, we hypothesize that these boundaries are the results of using some heuristic rules to make decisions in the task. To formulate decision making by these heuristic rules as a computational framework, we use the fuzzy logic theory. Based on this theory, we propose a new computational framework for decision making in evidence accumulation tasks. In this framework, there is no explicit decision boundary. Instead, the subject's desire to stop accumulating evidence and responding at each moment within a trial and for a given value of the accumulated evidence, is determined by a set of fuzzy ``IF-TEHN rules". We then use the back-propagation method to derive an algorithm for fitting the fuzzy model to each participant's data. We then investigate how the difference in the participants' performance in the experiment is reflected in the difference in the parameters of the fitted model
\end{abstract}


\end{frontmatter}
	
\section{Introduction}

Sequential sampling models have provided a rigorous computational framework for studying different decision making tasks (\cite{ratcliff_theory_1978,smith_psychophysically_1995,ratcliff_comparison_2004,palmer_effect_2005,kiani_bounded_2008,wagenmakers_diffusion_2008,busemeyer_decision_1993}). A common assumption among these models is that noisy evidence favoring each choice in the task is accumulated over the course of a trial and the subject responds as soon as the accumulated evidence favoring one of the choices reaches a \textit{decision boundary}. Different instantiations of these models differ in their assumptions about how the evidence is accumulated and how the decision boundaries are implemented (\cite{smith_stochastic_2000,usher_time_2001,brown_ballistic_2005,teodorescu_disentangling_2013,jones_unfalsifiability_2014,khodadadi_mimicry_2015}).

In many of the previous applications of these models, the decision boundaries have been assumed to remain constant during a trial. These models have been successfully fitted to the data from a wide range of tasks. The research on models with time-varying decision boundaries is less prevalent. A few studies have used these models to explain the data in the deferred decision making tasks (\cite{sanders_decision_1967,busemeyer_psychological_1988}). These models have been mostly investigated in the theoretical studies either as the optimal solution in specific experimental designs (\cite{rapoport_models_1971,ditterich_stochastic_2006,frazier_sequential_2008,drugowitsch_cost_2012,khodadadi_learning_2014}) or the solution of the model mimicry problems (\cite{zhang_time-varying_2014,khodadadi_mimicry_2015}). Recently, \cite{hawkins_revisiting_2015} and \cite{voskuilen_comparing_2016} compared time-constant and time-varying versions of the \textit{drift diffusion model} (a popular variant of the sequential sampling models) by fitting them to data from several behavioral experiments. Overall, their results showed that for the human data the additional complexity in the models with time-varying boundaries does not improve the quality of fit and therefore this additional complexity in unnecessary.

In contrast to these results, in a recent study, we found strong evidence supporting the models with time-varying decision boundary (\cite{khodadadi_learning_2016}). We argued that this inconsistency could be due to several factors: First, we used a novel stimulus (see the next section) which made it possible to observe the decision boundaries directly. Second, the subjects in our experiment were paid based on their performance. Third, in our experiment, by fixing the total time of the experiment, the subjects had to balance between their speed and accuracy to achieve the maximum reward. Fourth, in the computational models we considered there was a explicit mechanism for adjusting the decision boundary based on the feedback received in each trial (see the discussion section of \cite{khodadadi_learning_2016} for more details).  

The results of model comparison showed that the subjects adjusted their boundary during the experiment. Interestingly, using computational modeling of the subjects' data we found that the subjects used time-varying decision boundaries even at the beginning of the experiment and without any prior experience. This shows that the time-varying boundaries are not necessarily the results of the learning process. This raises the question that how do the subjects use time-varying boundaries even at the beginning of the experiment. In the previous studies of the time-varying decision boundaries, the boundaries were either modeled as a parametric function (Weibull function in \cite{hawkins_revisiting_2015,voskuilen_comparing_2016,khodadadi_learning_2016} and sum of basis functions in \cite{drugowitsch_cost_2012}) or as the solution obtained by dynamic programming (\cite{rapoport_models_1971,frazier_sequential_2008}). Representing the time-varying boundaries by a parametric function is useful because it makes it possible to fit the model to the data. However, it does not explain why the subjects use such boundaries. Deriving the time-varying boundaries as the solution of the optimality problems provides an explanation for why the subjects might use such boundaries. However, since the results in \cite{khodadadi_learning_2016} showed that the subject used these boundaries even at the beginning of the experiment, learning the optimal strategy could not be the only reason that the subjects use these boundaries.

Another explanation for time-varying boundaries, specifically time-decreasing boundaries, is provided by the \textit{urgency-signal models} (\cite{churchland_decision-making_2008,cisek_decisions_2009}). Based on these models, as time elapses in a trial, the subject needs less evidence to make her choice. This urgency to respond, can be interpreted as time-decreasing decision boundary. Several neurophysiological studies have found supporting evidence for this assumption. Although this framework provides an explanation for time-decreasing boundaries, it does not specify why people might use a particular form of the urgency signal. Therefore, to fit these models the urgency signal is modeled as a parametric function. For example \cite{cisek_decisions_2009} assumed that the urgency signal increases linearly with time, while \cite{ditterich_stochastic_2006} and \cite{hawkins_revisiting_2015} modeled it as a logistic function. Therefore, as before, these models do not explain why a particular form of the urgency signal is used.

In this paper, we take another approach for modeling time-varying decision boundaries. We conjecture that these boundaries are the results of using some heuristic if-then rules to make the decisions. We use the theory of \textit{fuzzy logic} to construct a computational model for decision making in an evidence accumulation decision making paradigm. Specifically, we model the decision making process as a \textit{fuzzy logic controller}. We then propose a method based on the \textit{back-propagation} algorithm for fitting the model to each subject's data. By fitting the fuzzy model we can investigate what heuristic rules each subject used throughout the experiment. Another advantage of this framework is that the subjects' prior knowledge is modeled explicitly by some \textit{linguistic variables} (see Section \ref{sec:fuzzy_model}). We finally, propose a reinforcement learning algorithm for adjusting the parameters of the fuzzy model after receiving feedback in each trial.

\section{Experiment}
\label{sec:experiment}
Data from this experiment were previously reported in \cite{khodadadi_learning_2016}. We used a novel stimulus in this experiment which we call \textit{canoe movement detection task}. At the beginning of each trial, a canoe was shown at the center of the screen. The canoe would move randomly back and forth to the left and right until the subject responded. The subjects were told that the canoe will eventually go to one of the two directions and were asked to respond by pressing the `m' or `c' key on the keyboard as soon as they decided that the canoe is going to right or left, respectively. The canoe movement was governed by a Markov chain which after each 500 msecs would jump to the correct direction with probability $P_0$ and the other direction with probability $1-P_0$. The difficulty of the trial could be controlled by the value of $P_0$: for larger values of this parameter, the canoe moved more coherently toward the correct direction and so the trial was easier. We recorded the time the subject responded (reaction time) as well as the canoe position (relative to the center of the screen) at the time the subject responded in each trial.

In this task, as the canoe gets closer to either ends of the screen, the probability of responding correctly increases. However, it takes longer for the canoe to reach positions further away from the center. Therefore, the participants could balance between their speed and accuracy by the position of the canoe at the time they responded. Thus, the position of the canoe at the time the subject responded in each trial can be considered as the value of the decision boundary at that time. This results in an advantage of this stimulus over the conventional stimuli used in evidence accumulation experiments (e.g., random dot motion task): the value of the decision boundary at the time the decision was made can be observed directly in this task. In contrast, in the previous tasks we could only observe the reaction time and the properties of the decision boundaries should be inferred indirectly, for example by using computational modeling.

We will use the data from Experiment A of \cite{khodadadi_learning_2016}. This experiment, consisted of 40 blocks of trials of the canoe movement detection task explained above. Each block was one minute long. Since the blocks' duration was fixed, the number of trials each subject could experience during each block (and during the whole experiment) depended on her speed. Each trial could come from one of two possible conditions: easy or hard. In the easy trials $P_0=0.65$ and the subjects would receive or lose 20 coins for their correct and incorrect responses. In the hard trials, $P_0=0.51$ and the pay-off was $\pm1$ coin. In addition, in the easy trials, after the incorrect responses the subject has to wait 3 more seconds before the next trial started. This delay penalty did not exist for the hard trials. 

To achieve the maximum number of coins, a subject must balance between her speed and accuracy. Specifically, she should spend much less time on the hard trials than the easy trials. The goal of the experiment was to investigate if the human subjects can learn the optimal strategy and how they adjust their decision boundary after receiving the feedback in each trial.
 
\section{Shape of decision boundaries}
In \cite{khodadadi_learning_2016} we investigated both the shape of the decision boundary that the subjects used in the experiment and how they adjusted the boundary after each trial. To this end, we developed and compared 10 computational models. The models differed in their assumptions on the shape of the decision boundary and the learning mechanism for adjusting it. Specifically, we consider 5 classes of learning mechanisms. For each class, we considered both time-constant and time-varying decision boundaries. The results of the Bayesian model comparison showed that for each class of models, the time-varying version fitted much better than the time-constant version. In addition, our model which would use a reinforcement learning algorithm to adjust the decision boundaries was the best model for most of the participants (Model 9 in \cite{khodadadi_learning_2016}). 

In this model the decision boundary was represented as a Weibull function (as was suggested by \cite{hawkins_revisiting_2015}) which has the following form: 

\begin{equation}
b(t)=\psi-\bigg[1-\exp\bigg(-\bigg(\frac{t}{\lambda} \bigg)^{\phi} \bigg) \bigg]\cdot \bigg[\frac{1}{2} \psi - \psi' \bigg]
\label{eq:WeibullThresh_01}
\end{equation}

In this equation, $b(t)$ is the value of the decision boundary for the ``right" response at time $t$. In this paper, we only consider symmetric boundaries and so the boundary for the left response is $-b(t)$. The free parameters of this model of the decision boundary are $(\psi, \psi',\lambda,\phi)$. The parameter $\psi$ was updated after each trial using a reinforcement learning algorithm. Suppose that in trial $k$ the reaction time was $t_k$ and the canoe position at the time the subject responded was $x_k$. We assumed $x_k=b_k(t_k)+\epsilon$, where $\epsilon\sim N(0,\sigma^2)$ is a zero-mean Gaussian random variable, and $b_k$ is the decision boundary in trial $k$.

\begin{figure}[!h]
	\centering
	\includegraphics[width=.7\linewidth]{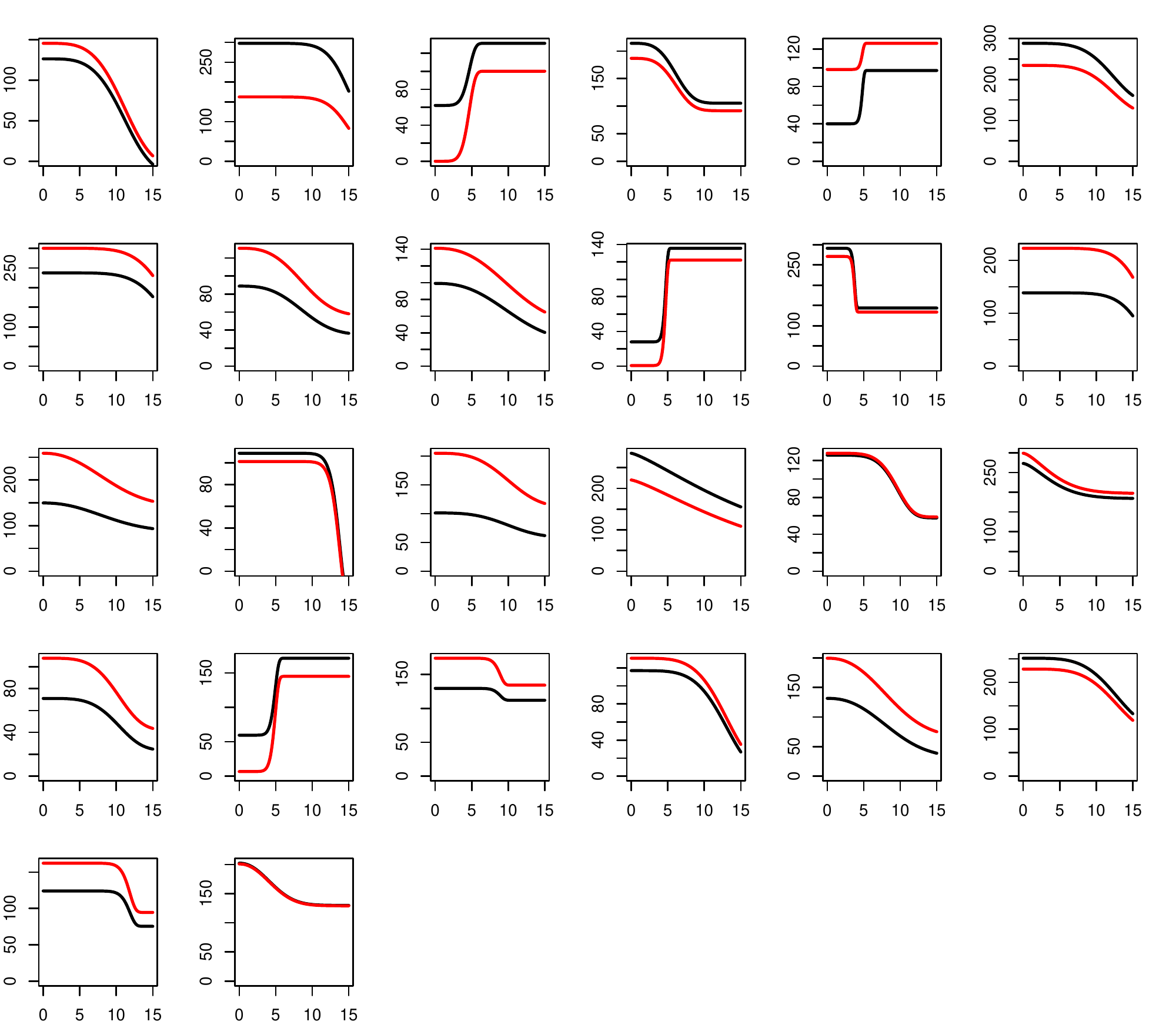}	
	\caption{{\bf Predicted boundaries}. Each panel shows the values of the decision boundaries used by a participant after the first (black curves) and last (red curves) block of trials in Experiment A of \cite{khodadadi_learning_2016}. The horizontal axis is elapsed time in a trial and the vertical axis is the canoe position. The boundaries are obtained by simulating the best fitted model. See the text for more details.} 
	\label{fig:Fitted_boundaries}
\end{figure}

After estimating the free parameters of the model for each subject, we can simulate the model with the fitted parameters to obtain the shape of the decision boundary in each trial. Figure \ref{fig:Fitted_boundaries} shows the decision boundary for the easy trials at the end of the first and last blocks of trials predicted by this model for each subject. As it can bee seen in this figure, most subjects used time-varying decision boundaries even in the first blocks of trials. These results raise the following question: how do the participants adopt these time-varying decision boundaries even at the beginning of the experiment and without any experience with the task?

The Weibull function is an appealing choice for describing the time-varying boundaries because it is a flexible function that can generate different patterns for the decision boundary with only a few parameters. The shape of the fitted boundaries modeled with this function, helped us to get a sense about the computational complexity needed to describe the decision mechanisms that the subjects used in the experiment. However, this does not provide a psychologically plausible explanation of these mechanisms. Specifically, the relationship between the values of the parameters of the Weibull function and its shape is complicated. The Weibull function (or any other parametric representation of the decision boundary) is useful for detecting any relationship between the value of the decision boundary and the elapsed time in a trial. However, it is not able to explain why such relationship exists. In other words, the parameters of this function are not psychologically interpretable.

\section{Accumulated evidence paths with zero-likelihood}
\label{sec:zeroLL}
\begin{figure}[!h]
	\centering
	\includegraphics[width=.7\linewidth]{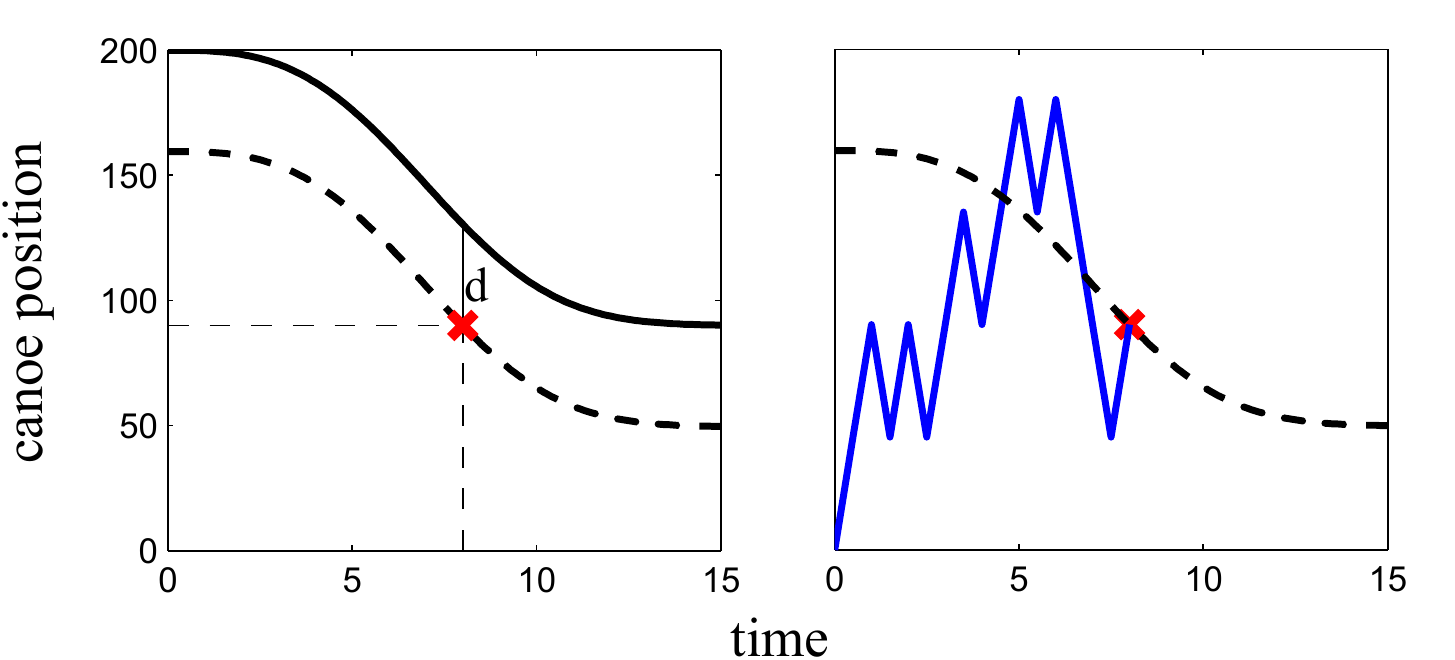}	
	\caption{{\bf Example of accumulated evidence path with zero likelihood}. Left: the data point $(t_k,x_k=(8,90))$ used in the example is shown as a red cross. The mean boundary $b(t)$ is shown in solid black. We have $d=b(t_k)-x_k$. Right: the accumulated path is also shown in this figure. The likelihood of the path is zero. See the text for more details.} 
	\label{fig:zeroLL}
\end{figure}

Besides the shape of the decision boundaries, we had another motivation to develop a new computational model for the canoe task. We explain this by an example. Suppose that we are trying to estimate the parameters of the Weibull function to the data of a participant using the maximum likelihood method. This is usually done using an optimization method which computes the likelihood of the data for several values of the parameters and finds the values for which the likelihood of the data is maximized. The first step in the maximum likelihood estimation is to decide what data we want to fit the model to. In most of the previous tasks for which the sequential sampling models have been used, the observable data in each trial is the subject's choice (correct or incorrect) and the reaction time. This is because in these tasks the decision boundary and the accumulated evidence are not observable. In the canoe task, however, both these are observable. For now, suppose that we want to ignore the accumulated evidence and fit the Weibull boundary to the choice, reaction time and the value of the decision boundary in each trial. As an example, suppose that in a trial the value of the subject's reaction time, $t_k$, and decision boundary, $x_k$, were 8 secs and 90 pixels. Also, suppose that we want to compute the likelihood of this data point for a Gaussian boundary (see the previous section) with mean and variance $b(t)$ and $\sigma^2$, where $b(t)$ is a Weibull function with parameters $(\psi, \psi',\lambda,\phi)=(200,50,5,2)$. The mean boundary and the data point are shown in the left panel of Figure \ref{fig:zeroLL}. To compute the likelihood of the data point, we assume that the boundary in the trial was a sample $y_k(t)$ from the Gaussian distribution such that $y_k(t_k)=x_k$. This boundary is shown with the dash line in the figure. It can be seen that the likelihood of the data point is $\exp[-d^2/(2\sigma^2)]/\sqrt{2\pi\sigma^2}$. This example shows how the likelihood can be computed when the observed data in each trial is just the canoe position and reaction time.

In the canoe task, in addition to the value of the decision boundary we can observe the canoe path as well. Suppose that the canoe path in the previous example was the blue curve shown in the right panel of Figure \ref{fig:zeroLL}. As it can be seen in the figure, for the given value of the Weibull function, this curve crosses the boundary before time $t_k=8$ secs. Therefore, the likelihood of observing this canoe path given those parameters is zero. During the maximum likelihood estimation of the parameters, the optimization routine might suggest parameter values which result in canoe paths with zero-likelihood in several trials. These make the logarithm of the likelihood unbounded which in turn causes instability in the estimates. To avoid this problem, in \cite{khodadadi_learning_2016} we ignored the canoe paths and fitted the models only to the reaction time and the decision boundaries data. 

This problem arises because the decision boundary partitions the tow-dimensional space of time and canoe position into two parts. In one part, the probability of responding is zero and in the other part the probability of responding is 1. As we will see, this problem does not exist in the fuzzy model because in this model the probability of responding at each point in this two-dimensional space is non-zero.

\section{A fuzzy logic model of time-varying decision boundaries}
\label{sec:fuzzy_model}
\begin{figure}[!h]
	\centering
	\includegraphics[width=.5\linewidth]{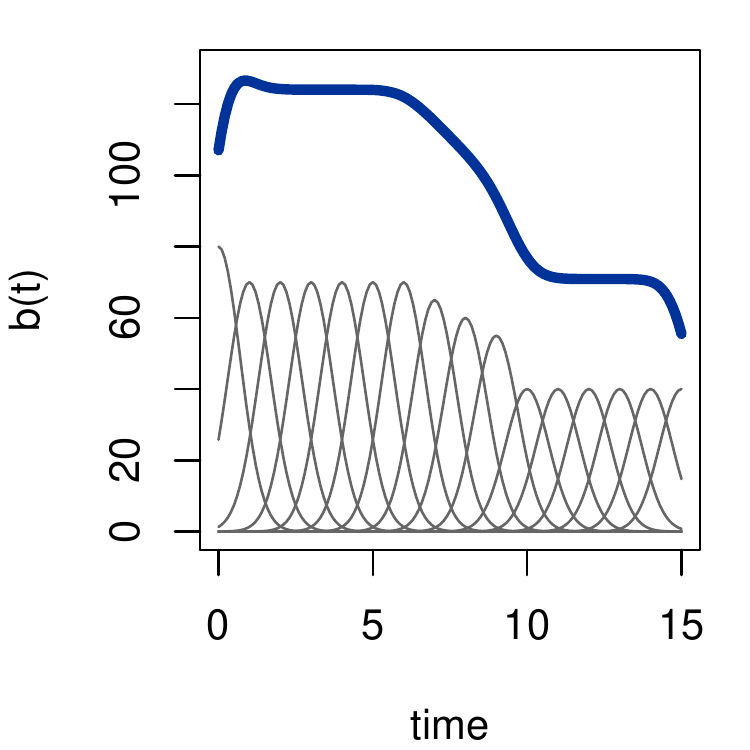}	
	\caption{{\bf Decision boundary as weighted sum of basis functions.} The boundary $b(t)$ is modeled as a weighted sum of basis functions. The weighted basis functions, $\omega_j \cdot B_j(t)$, are shown in dark gray and the boundary $b(t)$ is shown in dark blue.} 
	\label{fig:SumOfGausianThresh}
\end{figure}

Before explaining our model, we first consider another possibility for modeling the time-varying boundaries which is more plausible than the Weibull function. In this model, the boundary $b(t)$ is a weighted sum of a set of basis functions, that is: $b(t)=\sum_{j=1}^{n_b} \omega_j \cdot B_j(t)$, where $\omega_j$ are the weights and $B_j$ are the basis functions. Figure \ref{fig:SumOfGausianThresh} shows an example of $b(t)$ where the basis functions are all Gaussian functions. The relationship between the free parameters of this model of the time-varying boundary (the mean and variance of the Gaussian basis functions and the weights) and the shape of the resulting boundary is simpler than the Weibull function. For example if the weights are decreasing in time so will be the resulting boundary. However, this model still has the problem mentioned above: how does a subject sets the initial values of the weights?    

To address this problem we take another approach in modeling the time-varying boundaries. We model the boundary as the result of some heuristic rules that the subjects might use. At the beginning of the experiment, a subject's knowledge about how to perform in the experiment might be summarized as follows:

\begin{itemize}
	\item First, since there is a penalty for incorrect responses, she must not make many mistakes. To reduce the chance of an incorrect response, she must not make a decision before the canoe is ``far away enough" from the center of the screen.
	
	\item Second, since the time is limited, she should not spend ``much time"  on one trial. 
\end{itemize}

In other words, the subject only has a verbal description of how to perform in the experiment. More precisely, we conjecture that the subject's knowledge can be summarized as a set of verbal IF/THEN rules on two variables in each trial: the elapsed time and the canoe position. For example consider the two following rules:

\begin{itemize}
	\item \textit{IF the canoe is far from the center in a short amount of time THEN respond.}
	\item \textit{IF the canoe is not far from the center and not much time has elapsed THEN do not respond.}
\end{itemize}

Each rule consists of two parts: a \textit{premise} and a \textit{consequent}. The premise is defined on the subject's observation (the canoe position and the time) and the consequent is defined on the actions (or decisions) that the subject can take. These rules are not precise: they only provide an abstract description of what to do in each situation.

The question is how to turn this abstract description of the subject's knowledge into a formal computational model. The theory of \textit{fuzzy logic} provides the means to do this. In what follows, we show how the decision making in the experiment can be modeled as a \textit{fuzzy logic control} problem, and specifically how the time-varying decision boundaries can be considered as the output of a fuzzy logic controller. Our presentation of the fuzzy systems is mostly based on \cite{passino_fuzzy_1997}. 

The main component of a fuzzy controller is a \textit{rule-base}, a set of rules which specify what to do in each situation. The rules are defined on some \textit{input} and \textit{output} variables. The inputs determine the situation and the outputs determine the actions to take in each situation. For our experiment, the obvious choice for the inputs are the canoe position ($POS$) and the elapsed time ($T$) at each moment in a trial. The decision problem at each moment within a trial is if to stop accumulating evidence and respond or to continue accumulating evidence. For now, we define the output to be a continuous variable which encodes the subject's desire to respond at each moment.

\begin{figure}[!h]
	\centering
	\includegraphics[width=1\linewidth]{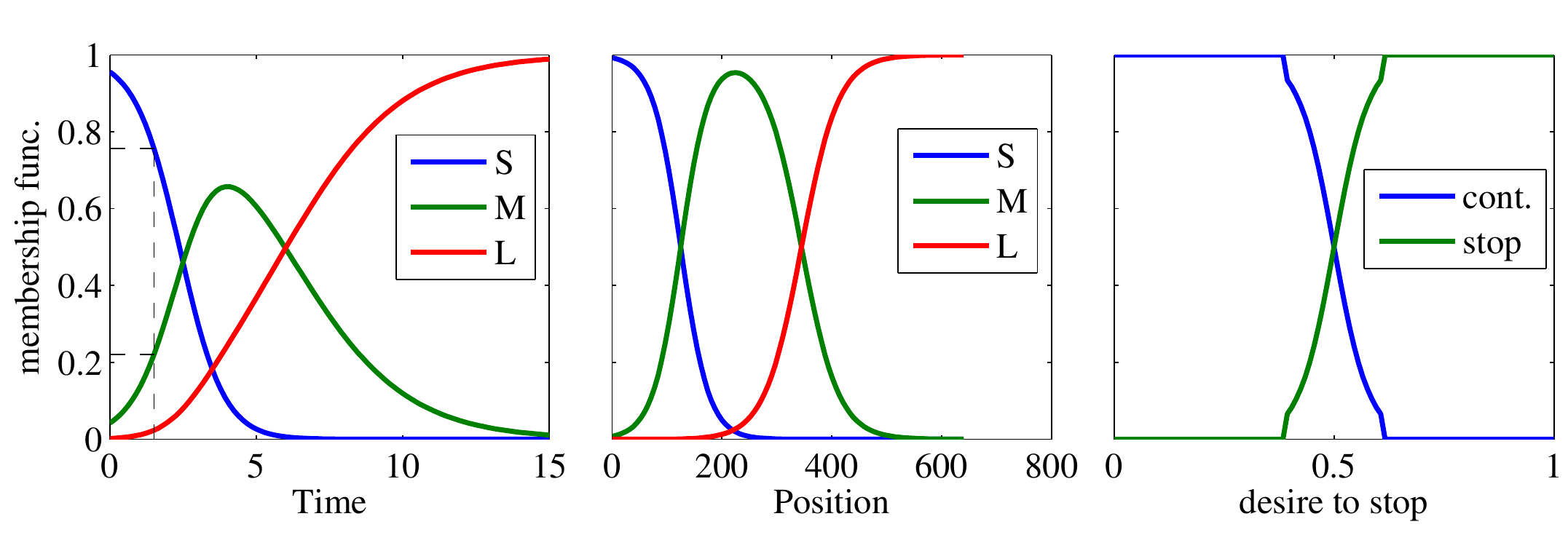}	
	\caption{{\bf Fuzzy membership functions for input and output variabls.} Left: membership functions corresponding to the labels ``short" (S), ``medium" (M) and ``long" (L) defined on the input variable $T$ (elapsed time in a trial). Middle: membership functions corresponding to the labels ``small" (S), ``medium" (M) and ``large" (L) defined on the input variable $POS$ (canoe position). Right: membership functions corresponding to the labels ``continue" (cont.) and ``stop"  defined on the output variable.} 
	\label{fig:MembershipFuncs}
\end{figure}

The rules in a fuzzy controller are defined on the verbal description of the inputs and outputs. Therefore, the first step in building a fuzzy controller is to assign some verbal labels to each variable. For example consider the elapsed time. We can describe each value of the elapsed time as being \textit{short}, \textit{medium} or \textit{long}. Now consider a rule which starts with ``IF time is short ...". Also suppose that 1.5 secs has elapsed since the beginning of a trial. Then the following question arises: should we consider 1.5 secs as short? It can be considered as short but it is also somewhat medium (and perhaps not so long). The theory of fuzzy logic allows to model this uncertainty about the mapping from the values of the variables and the verbal labels. This is done using \textit{membership functions}. Formally, a membership function is a mapping from the domain of the values that a variable can take to the interval $[0,1]$. It quantifies the certainty of the subject on how much each value of a variable belongs to (is a member of) a label. To clarify this, let us consider the label ``medium" for the input variable $T$. A possible membership function is shown in the left panel of Figure \ref{fig:MembershipFuncs} (together with the membership functions of the labels ``short" and ``long" for this variable). We will only consider the Gaussian membership functions which have the following form:

\begin{equation}
m_i(x)=\exp\bigg(-\frac{(x-\mu_i)^2}{\sigma_i^2}\bigg)
\label{eq:GaussianMember}
\end{equation}

\noindent where $x$ is the value of the input (time or position) and $\mu_i$ and $\sigma_i$ are the mean and standard deviation for the $i^{th}$ verbal label. In the figure, the parameters of the membership function corresponding to the label \textit{medium} for time are $\mu=5,\sigma=2$. The subject is ``0.22 certain" that 1.5 secs is \textit{medium}. Also, this subject is 0.76 certain that 1.5 secs is short. It is important to note that the membership functions are not probability distributions: they do not integrate to 1. A membership function only quantifies the degree of belief of a subject about how much a value belongs to a label. Figure \ref{fig:MembershipFuncs} shows examples of the membership functions defined on the input and output variables for modeling the subjects' decision making in the canoe Experiment. We assume that the subject uses the labels \textit{small}, \textit{medium}, and \textit{large} for the variable \textit{POS}, and the labels \textit{stop} and \textit{continue} for the the output variable. We have normalized the membership functions for each variable such that they sum to 1 for all values of a variable.

\begin{figure}[!h]
	\centering
	\includegraphics[width=.5\linewidth]{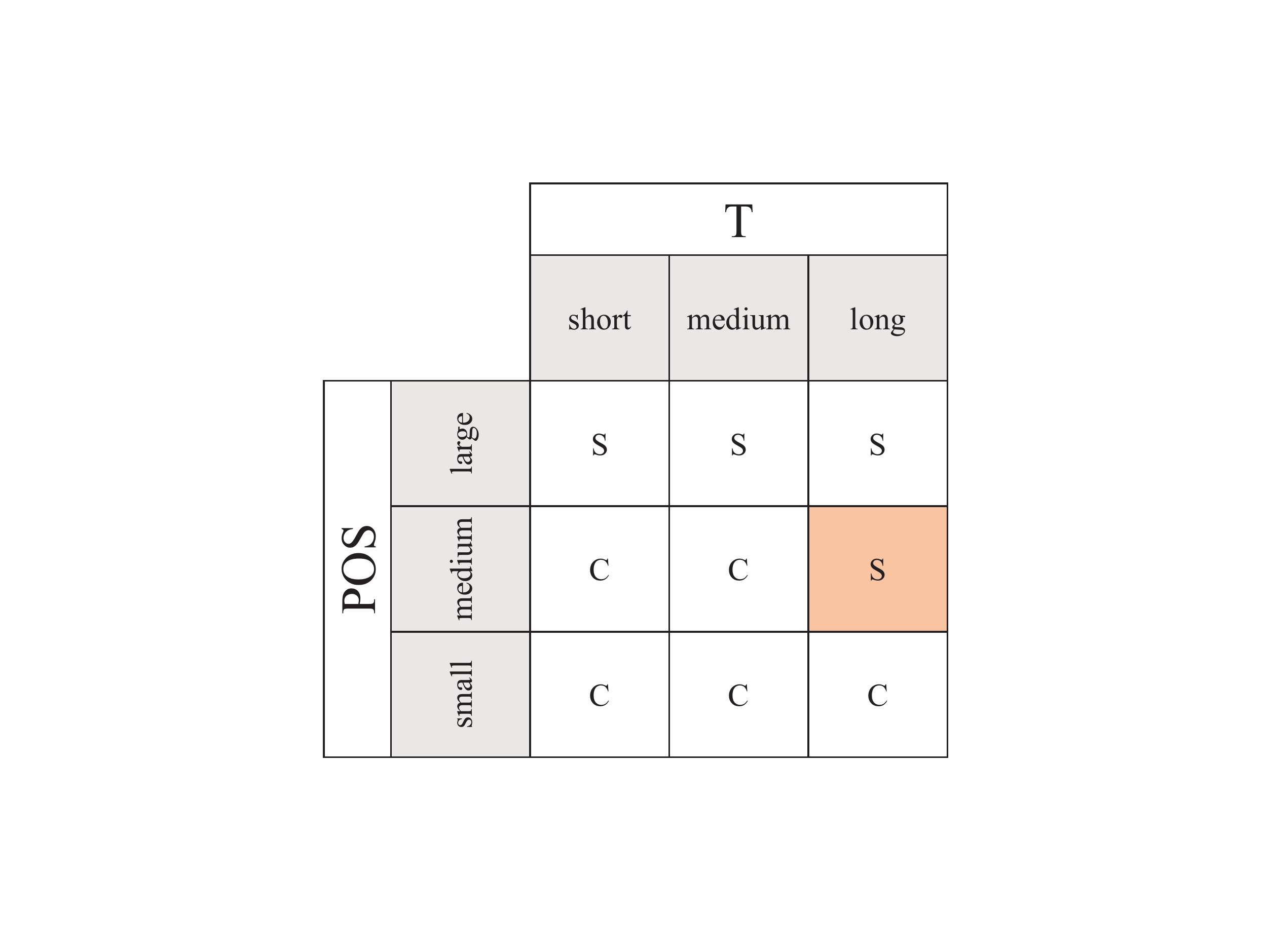}	
	\caption{{\bf Fuzzy rule-base.} The rows correspond to the labels of the input variable $POS$ and the columns correspond to the labels of the input variable $T$. The letter in each cell shows the desired label of the output variable. S:stop, C:continue. Each cell specifies one rule in the rule-base. For example the highlighted cell corresponds to the following rule: \textbf{IF} POS is medium \textbf{AND} T is long \textbf{THEN} stop.} 
	\label{fig:Fuzzy_RuleBase}
\end{figure}

After determining the labels for the variables, we should specify the rules acting on these labels. In other words, we should specify the rule-base. A convenient way to specify a rule-base is to use a tabular representation. One possible set of rules are shown in Figure \ref{fig:Fuzzy_RuleBase}. As we will see, this rule-base together with the membership functions shown in Figure \ref{fig:MembershipFuncs}, result in decision boundaries similar to those used by some of the subjects in the canoe Experiment (Figure \ref{fig:Fitted_boundaries}). It is easy to read the rules off this tabular representation. For example the highlighted cell represents the following rules:

\textit{\textbf{IF} POS is medium \textbf{AND} T is long \textbf{THEN} stop.}

The next step in building the fuzzy controller model is to specify how to evaluate each rule for given values of the input variables. To clarify this, consider the rule above. The premise of this rule consists of two parts: ``POS is medium'' and ``T is long''. Suppose that in a trial 1.5 secs has passed and the canoe position is 200 pixels away from the center. The question is how much this rule is valid for these values of the inputs. Using the membership functions shown in Figure \ref{fig:MembershipFuncs}, the subject is 0.76 certain that ``POS is medium'' and 0.22 certain that ``T is long''. To determine how much the premise of the rule is valid, we must find a way to combine the certainties on the two inputs. In other words, we need to quantify the logical ``AND" operation. In the theory of fuzzy logic, several simple methods have been proposed to do so. Here, we use the product rule: the amount of certainty of the subject for the premise of the rule is the product of the certainty to each of the components of the premise. Therefore, in the example mentioned above, the certainty to the premise of the rule is $0.76 \times 0.22=0.17$. For each value of the inputs, this quantity should be computed for the premise of all rules in the rule-base.

There are two more steps to complete the fuzzy controller model: First, computing the action that each rule recommends, and second, combining the recommendations from all rules and computing the value of the output for given values of the input. The comparison between different methods to perform these steps is outside the scope of this paper (for more details see for example \cite{passino_fuzzy_1997} Chapter 2). Here, we use simple methods for both steps. To compute the action recommended by a rule for given values of the inputs, we multiply the membership function corresponding to the consequent of the rule by the value of the premise of the rule. For example, consider the rule mentioned above again. The consequent of this rule is ``stop". The membership function for this action is shown in the right panel of Figure \ref{fig:MembershipFuncs} in green. To compute the recommended action by this rule, we simply multiply this membership function by 0.17, the value of the premise of this rule when the inputs are 1.5 secs and 200 pixels. The results of this step, is a scaled membership function for each rule. Next, we should combine these recommendations and compute a single output value for each given value of the inputs. We perform this step using a method called ``center-average defuzzification". Let $c_j$ denote the center of the membership function of the consequent of the $j^{th}$ rule. For example, the consequent of the above rule is ``stop" and, as it can be seen in the right panel of Figure \ref{fig:MembershipFuncs}, the center of the membership function corresponding to ``stop" is 0. Also, for given values of the inputs, let $y_{2,j}$ denote the premise value of the the $j^{th}$ rule (the reason for using these subscripts will be clear in the next section). Then, based on the center-average defuzzification method, the output is computed as follows:

\begin{equation}
	O=\frac{\sum_j y_{2,j} \cdot c_j}{\sum_j y_{2,j}}
	\label{eq:fuzzy_O}
\end{equation}

\begin{figure}[!h]
	\centering
	\includegraphics[width=.5\linewidth]{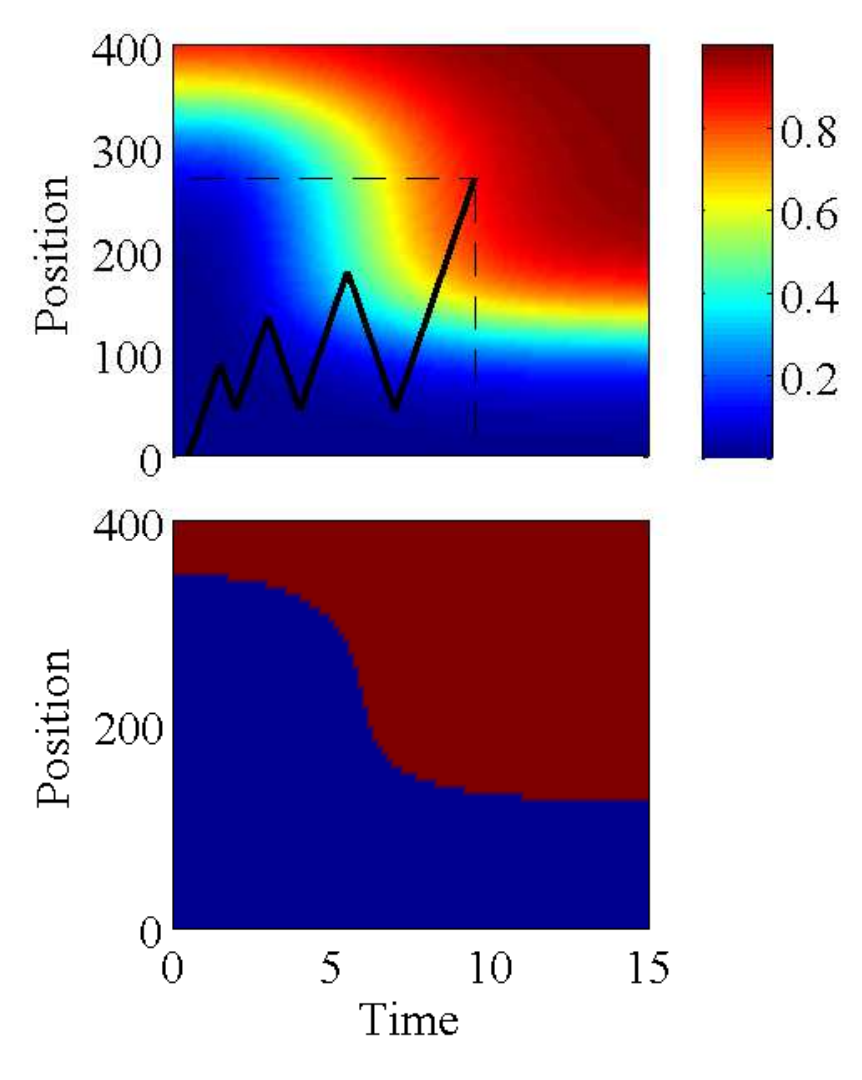}	
	\caption{{\bf Output of fuzzy model.} Top: The output $O$ of the fuzzy model computed using Equation \ref{eq:fuzzy_O} for a range of values of the elapsed time and the canoe position. Bottom: this figure is obtained by passing $O$ in the top panel through a boundary value of 0.5. The blue region corresponds to values of the inputs for which the subject continues accumulating information . The red region specifies the values for which the subject stops accumulating information and responds. The boundary between the two regions can be considered as the time-varying decision boundary.} 
	\label{fig:Fuzzy_Output}
\end{figure}

This completes the fuzzy controller model. The output $O$ of the fuzzy controller for the rule-base of Figure \ref{fig:Fuzzy_RuleBase} and the membership functions of Figure \ref{fig:MembershipFuncs} is shown as a heat map in the top panel of Figure \ref{fig:Fuzzy_Output} (the black curve is a simulate canoe path explained below). These values encode the subject's desire to stop accumulating information and responding for each value of the elapsed time and the canoe position in a trial. We can pass these values through a threshold value to come up with a single boundary function. The bottom panel of Figure \ref{fig:Fuzzy_Output} shows the results of this thresholding with threshold being equal to 0.5. The boundary between the two colored regions is the decision boundary.

The two panels of Figure \ref{fig:Fuzzy_Output} imply two different mechanisms for decision making in the canoe experiment. In the bottom panel, a single decision boundary is produced and the subject makes her decision when the canoe position reaches this boundary. This is the conventional model for the decision making in evidence accumulation experiments: the subject accumulates evidence favoring each choice and responds as soon as the accumulated evidence reaches the decision boundary. In the top panel of this figure, however, no decision boundary is produced. Instead, the function shown in this panel represents the subject's desire to respond at each point in the two-dimensional space of time and canoe position. We assume that this function is proportional to the probability of responding at each point. Let $p(t,x)$ denote the probability of responding at time $t$ and when the canoe position is $x$. Then we have $p(t,x)=\frac{O(t,x)}{o}$, where $O(t,x)$ is the output of the fuzzy system and $o$ is a constant. The black curve in the top panel of Figure \ref{fig:Fuzzy_Output} shows a simulated canoe path. In this simulated trial the subject has responded at 9.5 secs when the canoe was 270 pixels away from the center. We can compute the likelihood of observing this path. For the sake of simplicity, instead of computing the likelihood of the continuous path, we explain how one can compute the likelihood of discrete points on the path. Some discrete points on the path are indicated by crosses. Let $(t_1,x_1),...,(t_N,x_N)$ denote these points, where $(t_N,x_N)$ is the point at which the subject responded. The likelihood of these points based on the fuzzy model is:

\begin{equation}
p(t_N,x_N) \cdot \prod_{i=1}^{N-1} 1-p(t_i,x_i)
\label{eq:Fuzzy_likelihood}
\end{equation} 

In words, this likelihood is equal to the probability of not responding in all points before $t_N$ and then responding at $(t_N,x_N)$. This likelihood can be used to estimate the parameters of the model for the observed data of each subject (see the next section). 

Before explaining how we can fit the fuzzy model to the data, we return to the problem that was explained in Section \ref{sec:zeroLL}. In the fuzzy model the problem of zero likelihood canoe paths does not exist, because in this model (as it can be seen in the top panel of Figure \ref{fig:Fuzzy_Output}) the probability of responding at each point in the two-dimensional space of time and canoe position is non-zero. In the conventional sequential sampling models this space is partitioned into two parts. These parts are shown with two colors in the bottom panel of Figure \ref{fig:Fuzzy_Output}. In the blue part, the probability of responding is zero and the red part this probability is 1. As we explained in Section \ref{sec:zeroLL}, this might results in paths with zero likelihood.

\section{Fitting fuzzy model to data}
\label{sec:fit_fuzzy}
\begin{figure}[!h]
	\centering
	\includegraphics[width=1\linewidth]{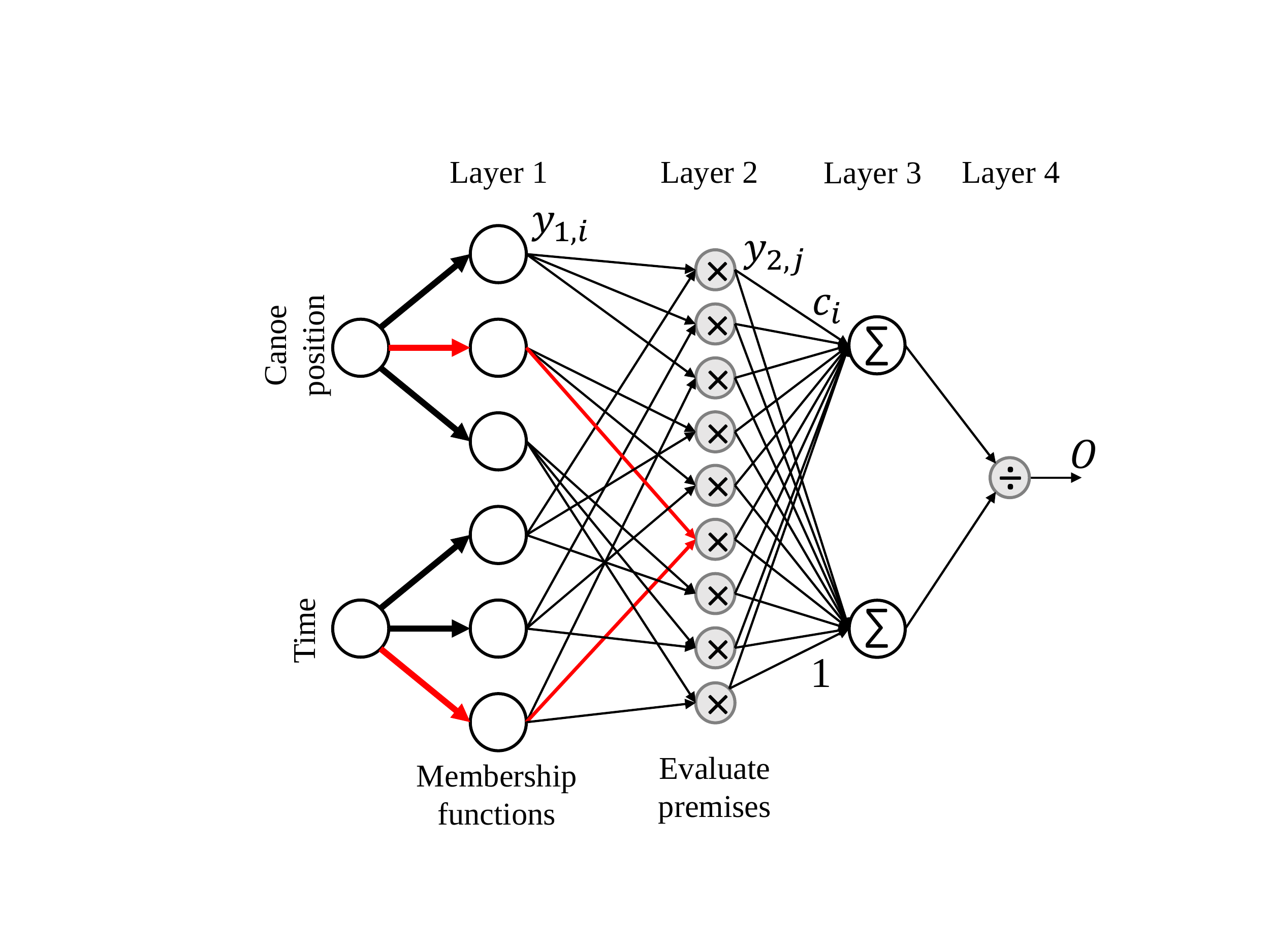}	
	\caption{{\bf Multi-layer neural network implementation of the fuzzy model.} This figure shows how the computations involved in the fuzzy model can be carried out using a multi-layer neural network. The first layer, implements the membership functions. The second layer evaluates the premise of all rules in the rule-base. The third layer computes the numerator and denominator of Equation \ref{eq:fuzzy_O}. See the text for more details.} 
	\label{fig:Fuzzy_NN}
\end{figure}

The computations performed in the fuzzy controller model can be carried out using a multi-layer neural network architecture. This architecture is shown in Figure \ref{fig:Fuzzy_NN}. This neural network consists of four layers. The input layer is connected to the neurons in the first layer. Each neuron in this layer has a membership function as its ``activation function". Since we considered 3 verbal labels for each input, there are the total of 6 neurons in this layer. The output of each neuron of this layer, $y_{1,i}$, is the certainty for each label of the inputs. The second layer evaluates the premises of the rules in the rule-base. Each neuron in this layer receives two inputs and multiplies them. This is equivalent to the product rule for premise evaluation. The number of neurons in this layer is equal to the number of the rules in the rule-base. The weights connecting layer 1 to layer 2 are all equal to 1. The output of each neuron of layer 2, $y_{2,j}$, goes to the third layer which has two neurons. The inputs to the upper neuron of the third layer are $y_{2,j}\cdot c_j$ and the inputs to the lower neuron are $y_{2,j}$. The neurons in layer 3 compute the sum their inputs. Layer 4 has only one neuron in which the output of the upper neuron of layer 3, $\sum_j y_{2,j}\cdot c_j$, is divided by the output of the lower neuron of layer 3, $\sum_j y_{2,j}$, which yields the output in Equation \ref{eq:fuzzy_O}. As an example, the path involved in computing the premise value for the rule ``\textit{\textbf{IF} POS is medium \textbf{AND} T is long \textbf{THEN} stop}'' is highlighted in Figure \ref{fig:Fuzzy_NN}.

Our goal here is to develop an algorithm to specify this architecture for each subject. Specifying this architecture consists of two major learning problems: \textit{structure learning} and \textit{parameter estimation} (\cite{kim_hyfis:_1999}). Structure learning is the process of identifying the fuzzy rules that the subjects has used (the rule-base). Parameter estimation is the process of estimating the free parameters of the network (the mean and variance of the membership functions). For structure learning we use the method proposed by \cite{wang_generating_1992}. For parameter estimation we use a gradient-descent algorithm.

Before explaining the fitting algorithms, we must specify the data that is used for model fitting. In the canoe experiment, we recorded the reaction time and the canoe position at the decision time in each trial. In addition, we could record the canoe position at each moment from the beginning of the trial until he subject responded in each trial. Due to some technical problems, we only recorded the canoe position at all moments for only two of the participants. We first explain the procedure of fitting when we have access to the canoe positions at all moments (and so the accumulated evidence path) and then explain the procedure for when we only have the reaction time and the canoe position at the decision time. 

\begin{figure}[!h]
	\centering
	\includegraphics[width=.5\linewidth]{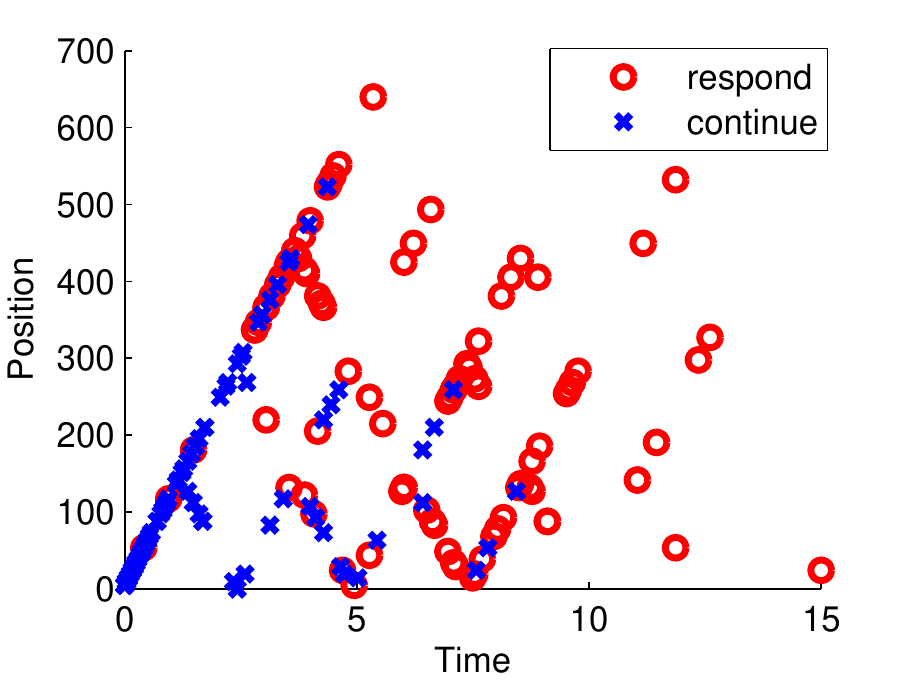}	
	\caption{{\bf Data for a sample subject in the canoe experiment.} The red circles are the points at which the subject has responded and the blue crosses are a random sample from the points that the subject continued accumulating information. See the text for more details.} 
	\label{fig:Sample_data_2class}
\end{figure}

In the previous section, we showed how one can compute the likelihood of observing the discrete points on the canoe path for each trial. There is however a problem in using this likelihood function as the loss function for model fitting: in each trial there is only one point that corresponds to responding (i.e., $(t_N,x_N)$) and all other points correspond to accumulating evidence. This may bias the estimated probability of responding toward zero. To address this problem, for each trial we take only one point on the canoe path randomly. Therefore, for each trial we have two observed data points: one at which the subject has continued accumulating evidence and one at which the subject has responded. These points for a sample subject are shown in Figure \ref{fig:Sample_data_2class}. In this way, the model fitting turns into a two-class classification problem. The data points belong to either ``continue" of ``respond" classes and the output of the fuzzy model specifies the probability that each point belongs to either class. Each data point is in the form $(t_i,x_i,y_i)$ where $y_i=0$ if $(t_i,x_i)$ belong to ``continue" and $y_i=1$ otherwise. The log likelihood of the observed data for a subject will be:

\begin{equation}
L=\sum_i [y_i \cdot \log(p(t_i,x_i))+(1-y_i) \cdot \log(1-p(t_i,x_i))]
\label{eq:cross_entropy}
\end{equation}

\noindent This is usually called the \textit{cross-entropy loss function} in the machine learning literature.

\subsection{Structure learning}
Next, we explain Wang and Mendel's algorithm for structure learning. The goal is to extract a set of fuzzy rules from the data. The algorithm consists of three steps.

\begin{figure}[!h]
	\centering
	\includegraphics[width=.5\linewidth]{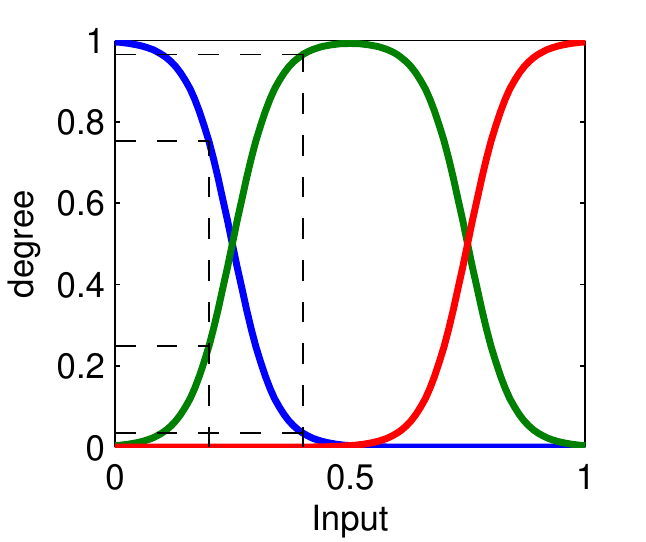}	
	\caption{{\bf Example of membership functions used for structure learning.} In the example explained in the text, we used the same membership functions for both input variables. In the figure, the degree that the input values (0.2,0.4) belong to each linguistic variable are shown. See the text for more details.} 
	\label{fig:StructureLearning_MemFunc}
\end{figure}

\textit{Step 1.} In this step, we first choose the number of the linguistic labels for input and output variables\footnote{The number of labels is a hyper-parameter of the model and can be determined by for example cross-validation.}. Then, the membership functions are set such that their centers have equal distance from each other and they cover the whole range of each variable. Although not necessary, we normalized all variables so their range is [0,1]. Three Gaussian membership functions used for the input variables are shown in Figure \ref{fig:StructureLearning_MemFunc}. For simplicity, we use the same membership functions with the labels \{S,M,L\} for both input variables. The membership functions for the output variable were shown in the right panel of Figure \ref{fig:MembershipFuncs}.

\textit{Step 2.} In this step one rule is generated for each data point. After normalization, each data point in our problem has the form $(t_i,x_i,y_i)$, where $t_i,x_i\in[0,1]$ and $y_i\in\{0,1\}$. To generate a rule for each data point, we first determine how much each variable of that data point belongs to the corresponding labels. For example consider the data point (0.2,0.4,1). As it can be seen in Figure \ref{fig:StructureLearning_MemFunc} the degree of 0.2 is 0.75, 0.25 and 0 for S, M and L labels, respectively. Similarly, the degree of 0.4 is 0.04, 0.96 and 0 for these labels. Degree of the output variable is 1 to the label `stop'. The rule is obtained simply by finding the maximum degree for each variable. Therefore, the rule corresponding to the data point (0.2,0.4,1) will be:

\textit{IF time is S AND position is M THEN stop.}

\textit{Step 3.} If there are $N$ data points in the training set, step 2 generates $N$ rules. Many of these rules may have conflict with each other. Specifically, some of the rules may have the same premise but different consequent. To resolve this conflict, we compute a \textit{certainty degree} for each rule. We compute this degree for each rule simply by multiplying the degree of each of its variables. For example for the rule generated in the previous step the degree of 0.2 to S is 0.75, the degree of 0.4 to M is 0.96 and the degree of the consequent is 1. The certainty degree of this rule is $0.75\times 0.96 \times 1 = 0.72$. Finally, among all rules with the same premise, we chose the one that has the highest certainty degree. The output of these three steps will be a rule-base without any conflict between the rules.

\subsection{Parameter estimation using gradient descent}
\label{sec:param_esimate}
The result of the structure learning algorithm is a fuzzy model in which the center of the membership functions are equidistant from each other. In addition, the variance of the Gaussian membership function is not chosen based on any optimality criteria. Therefore, the resulting model does not fit well to the data and we need to adjust the parameters of the model (the mean and variance of the Gaussian membership functions of the input variables) based on the observed data from each participant. As we showed, the fuzzy model can be represented as a feed-forward neural network. The most popular methods for estimating the parameters of such networks are based on the gradient descent algorithm and specifically the \textit{back-propagation of the error}.

Let $\theta_i$ denote a free parameter of the network. Suppose that the goal is to adjust the parameters such that a loss function, for example the cross-entropy $L$ in Equation \ref{eq:cross_entropy} is minimized. Based on the gradient descent method the parameter is updated as follows:

\begin{equation}
\theta_i^{new}=\theta_i^{old} - \eta \cdot \frac{\partial L}{\partial \theta_i}
\end{equation} 

\noindent where $\eta$ is the learning rate. Therefore, to use this method we need to evaluate $\frac{\partial L}{\partial \theta_i}$ for all $\theta_i$. For a feed-forward network, this is done through the back-propagation of error. In the Appendix, we have derived the formula for $\frac{\partial L}{\partial \theta_i}$. The free parameters are the center of the membership functions $\mu_i$ and the variance $\sigma_i^2$.

\section{Results}
In this section, we first present the results of fitting the fuzzy model to the data of the subjects who participated in three sessions of the experiment and we recorded the canoe path during each trial, and then present the results of fitting for subjects for which we do not have the canoe paths. Our goal here is not to make any conclusion about the data. Instead, we try to show how the patterns observed in the data are explained in the fuzzy modeling framework in terms of the fuzzy labeling of the input variables and the fuzzy heuristic rules.

\begin{figure}[!h]
	\centering
	\includegraphics[width=1\linewidth]{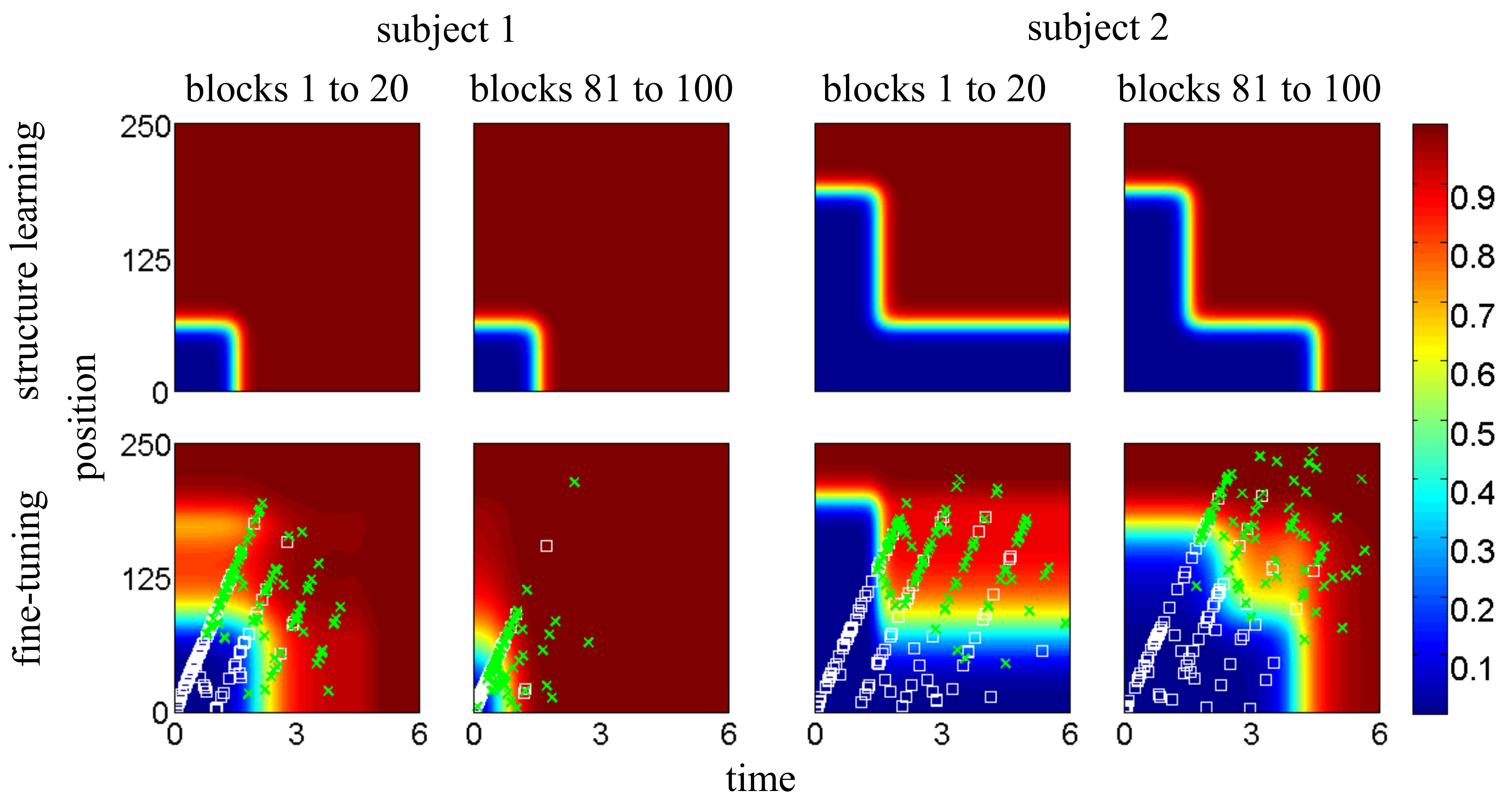}	
	\caption{{\bf Output of fitted fuzzy model.} The fuzzy model was fitted to the subjects participated in the 3 sessions of the canoe experiment. We fitted model to the data of both subjects, for the first and last 20 blocks separately. The top row, shows the output of the model after the structure learning stage, and the bottom row shows the output after fine-tuning the parameters using the back-propagation method. In the bottom row, the green crosses are the point at which the subject has responded in each trial while the white squares are one random sample from the canoe path in each trial.} 
	\label{fig:3session_fitted_output}
\end{figure}

\begin{figure}[!h]
	\centering
	\includegraphics[width=.5\linewidth]{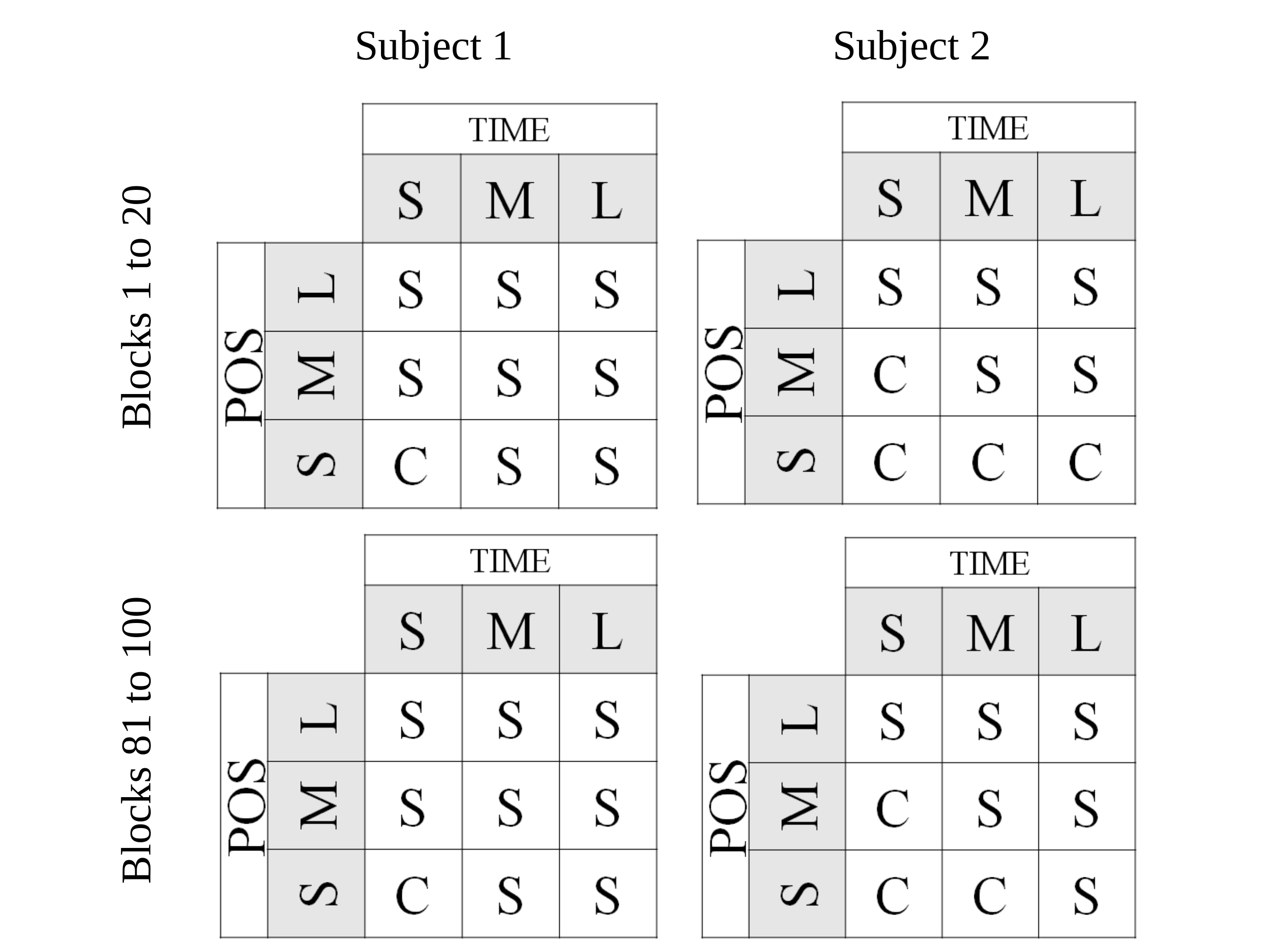}	
	\caption{{\bf Rule-based of fitted models.} Each table shows the rules extracted from the structure learning stage.} 
	\label{fig:3session_fitted_rulebase}
\end{figure}

\begin{figure}[!h]
	\centering
	\includegraphics[width=1\linewidth]{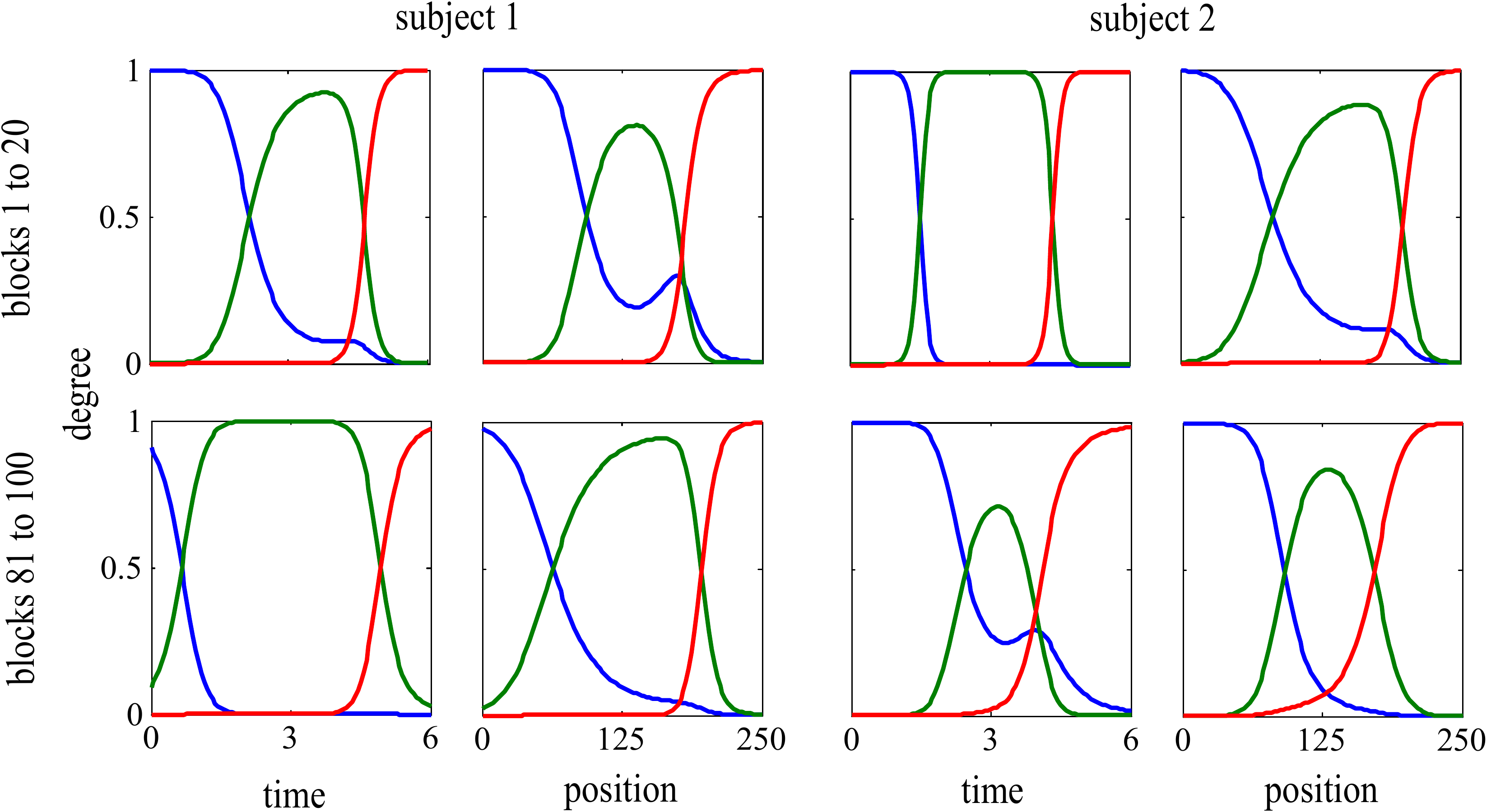}	
	\caption{{\bf Memberhsip functions of fitted models.} This figure shows the membership functions for the two input variables. The parameters are estimated for each subject using the back-propagation algorithm presented in the Appendix. This figure shows the difference between the two subjects in assigning each value of the inputs to the linguistic variables.} 
	\label{fig:3session_fitted_membership}
\end{figure}

Figures \ref{fig:3session_fitted_output}, \ref{fig:3session_fitted_rulebase} and \ref{fig:3session_fitted_membership} show the results of fitting the model to the data of the subjects in the 3-session version of the experiment. For each subject, we fitted the models to the data from blocks 1 to 20 and 81 to 100 separately. Figure \ref{fig:3session_fitted_output} shows the output of the fitted fuzzy model for a range of the input variables. The top row in this figure is the output after the structure learning stage and before fine-tuning the parameters of the membership functions. The bottom row shows the output after fine-tuning the parameters. The rules extracted by structure learning are presented in the rule-base tables of Figure \ref{fig:3session_fitted_rulebase}. Finally, the membership functions for the input variables (time and position) after fine-tuning the parameters are shown in Figure \ref{fig:3session_fitted_membership}.

As it can be seen in Figures \ref{fig:3session_fitted_output} and \ref{fig:3session_fitted_rulebase}, the two subjects have used different rules in the first 20 blocks of trials: the premise of the rule-base for subject 1 is `C' (continue) only if both time and position are `S', while for subject 2 the premise of four of the rules is `C'. This has led to dramatic difference between the shape of the output for the two subjects. Interestingly, the pattern of the output for blocks 81 to 100 (session 3) is similar for each subject to the pattern in blocks 1 to 20 (session 1). In other words, although the shape of the output has changed due to learning, the final shape resembles the initial shape of the output. This point is important because it shows that (at least for these two subjects) the performance during the experiment is affected a lot by the initial heuristic rules that a subject uses. 

Using Figure \ref{fig:3session_fitted_membership} we can investigate how the subjects labeling of the input variables has changed by practice. For example for subject 1, the range of the membership function corresponding to the label `S' for the time has decreased by practice while the range of the `M' membership function has increased. Based on Figure \ref{fig:3session_fitted_rulebase} we see that the rule-base for this subject has not changed from the first 20 blocks to the last 20 blocks. Therefore, for this subject, the change in the output of the fuzzy model from the first to the last blocks is not due to a change in the rule-base, and instead, this change is due to the change in the labeling of the input variables.
\newline

\begin{figure}[!h]
	\centering
	\includegraphics[width=1\linewidth]{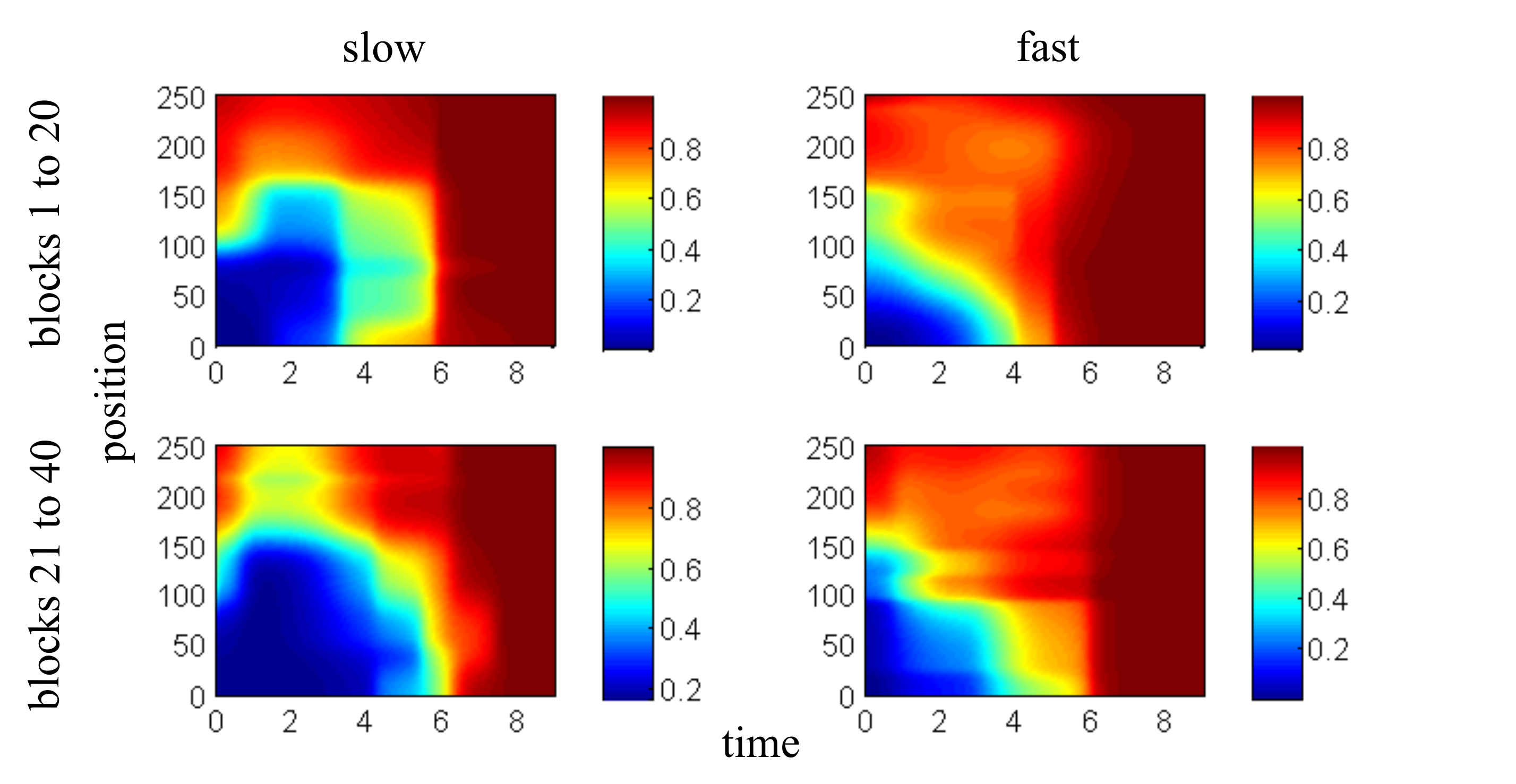}	
	\caption{{\bf Average of the output of fitted fuzzy model.} The left column is the average of the output for the slow group and the right column is the same values for the fast group participated in one session of the canoe experiment. See the text for how these values are computed and how the subjects are assigned to each group. The top and bottom rows show the results for the first and last 20 blocks, respectively. Notice that for the slow group, although the blue area has grown from the first to the last 20 blocks, but the minimum value is about 0.2 in the bottom panel while it is about 0 in the top panel. Therefore, the slow group have become faster by practice.} 
	\label{fig:1session_output}
\end{figure}

Next, we turn to analyzing the data of the subjects who participated in one session of the experiment. Here, we analyze the data of 26 subjects. Our goal here is to compare the properties of the fitted fuzzy models for the fastest and slowest subjects. To recognize these two groups, we first computed the median of the decision boundary (canoe position at the time the subject responded in each trial) in the first 20 blocks for each subject. Then we computed the 0.25 and 0.75 quantiles of these values. The subjects whose median decision boundary was less than 0.25 quantile were assigned to the fast group and the subjects whose median decision boundary was larger than 0.75 quantile were assigned to the slow group. There were the total of 7 subjects in the fast group and 6 subjects in the slow group. Next, we fitted the fuzzy model to the data from the first and last 20 blocks for each subject in each group. Then, we computed the output of the fitted model for each subject for a range of the input variables. The average of these values for each group and separated for the first and last 20 blocks is shown in Figure \ref{fig:1session_output}. As we mentioned before, the output of the fuzzy model is proportional to the probability of stopping at each point. It can be seen in the figure that the average output for the fast group is larger than the slow group. To investigate this difference more, we compare the rule-base used by each group of subjects. To this end, for each possible premise we compute the average of the consequent for each group of subjects. For example consider the following premise:

``IF time is M and position is S"

\begin{figure}[!h]
	\centering
	\includegraphics[width=1\linewidth]{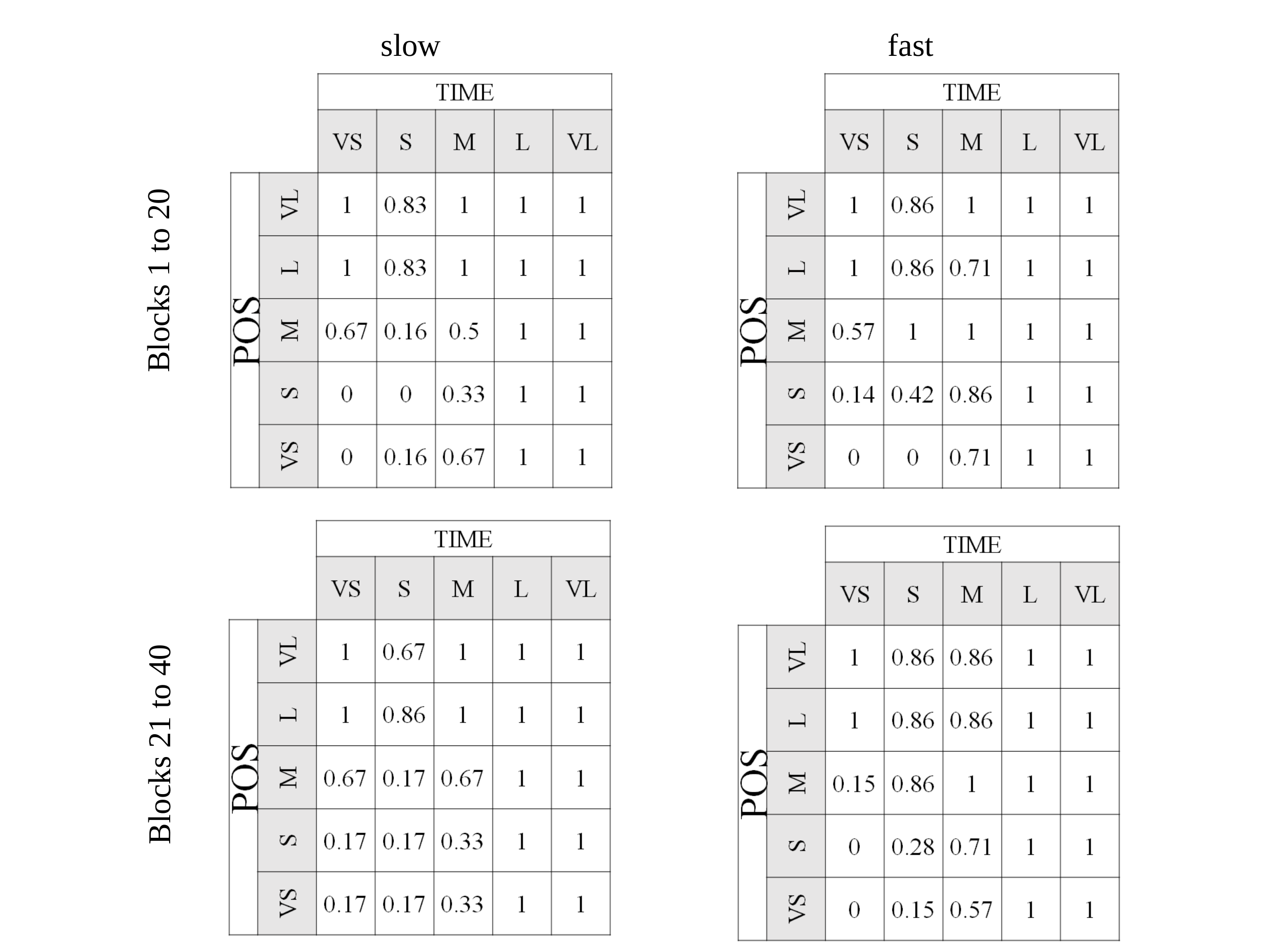}	
	\caption{{\bf Average of the consequents of rules in fitted fuzzy model.} Each cell in each table shows the value of the consequent of the corresponding rule in the fitted model, averaged across all subjects in each group. To compute these values, the consequent `continue' has been mapped to the value 0 and the consequent `Stop' has been mapped to 1. Therefore, higher values indicates higher desire to stop for a rule. As it can be seen, the average value of the cells in the bottom-left corner has increased for the slow group from the first to the last 20 blocks. This shows that these groups have become faster by practice. The reverse pattern can be observed for the fast group.} 
	\label{fig:1session_rulebase}
\end{figure}

The consequent of this premise for the 6 subjects in the slow group and for the first 20 blocks were [1,0,1,0,0,0], where 0 and 1 encode ``continue" and ``stop", respectively. Therefore, the average value of the consequent for this premise is 0.33 for this group. These values for all premises are shown in Figure  \ref{fig:1session_rulebase} separated by group and block number. In fitting the fuzzy model we have used 5 linguistic labels for each input variable. As it can be seen in this figure, the main difference between the two groups is the average value of the consequent when the time and position are either `S' or `M'. This average value is much higher for the fast group. Specifically, for the premise ``time is S and pos is M", all fast subjects have used the consequent ``stop" (the value of this cell is 1) while most of the slow subject have used continue (the value of this cell is 0.16).

\begin{figure}[!h]
	\centering
	\includegraphics[width=.4\linewidth]{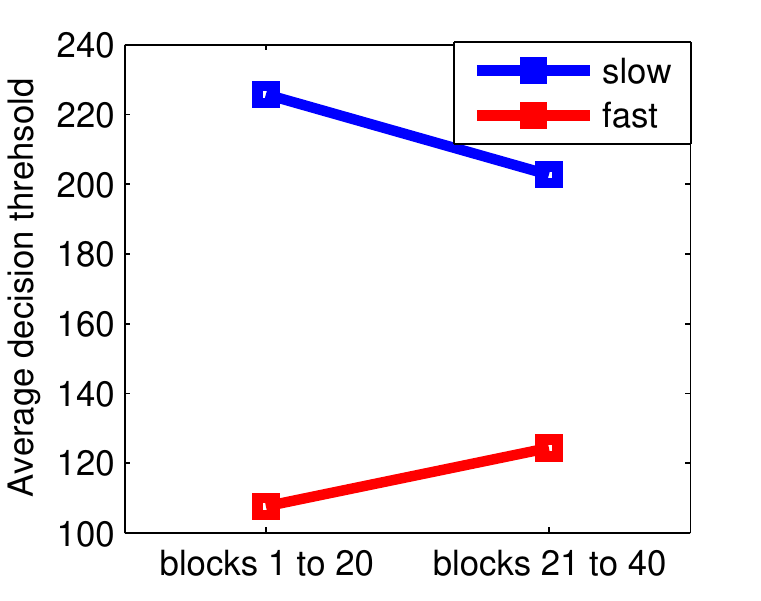}	
	\caption{{\bf Aveerage decision threshol.} The average value of the decision boundary for the two groups and for the first and last 20 blocks. As it be seen, the slow group have become faster by practice while the slow group have become slower.} 
	\label{fig:1session_averageboundary}
\end{figure}

Another interesting pattern emerges from the comparison of the average consequent values for the first and last 20 blocks for the two groups. For the slow subjects, the average values of the cell in the lower left corner of the table has increased (except for the premise ``if time is M and pos is VS") from the first to the last 20 blocks. For the fast subject, on the other hand, these values have decreased. This means that by learning, the slow subjects have become faster and fast subjects have become slower. The average value of the decision boundary for the two groups, separated by block number, is shown in Figure \ref{fig:1session_averageboundary}. Similar pattern can be observed in this figure. The average decision boundary is significantly higher for the slow group than the fast group in both first and last 20 blocks ($p<0.005$). However, the difference between the average value for the first and last blocks is not significant for either of the groups (for the fast subjects $p=0.18$, and for the slow subjects $p=0.29$). This might be due to the small sample size. 

Our goal in this section was to demonstrate how the fuzzy model can be used to interpret differences observed between subjects' performance. Using this model, we can attribute the observed patterns in the subjects' performance to the heuristic rules and/or the way each subject has assigned an input value to the corresponding linguistic labels.  

\section{A reinforcement learning algorithm for adjusting rules' consequents}
The fuzzy model enables us to explain the shape of the decision boundaries subjects used in the canoe experiment. However, it lacks an important aspect of the subjects behavior: the ability to adjust the decision boundary based on the feedback received in each trial. The results of model comparison in \cite{khodadadi_learning_2016} showed that for most of the subjects the model in which the boundaries were adjusted using a reinforcement learning algorithm provided the best fits. Specifically, this model performed much better than a model in which the boundaries would remain the same during the experiment (i.e., a model with no learning mechanism). In this section, we propose a reinforcement learning algorithm for adjusting the parameters of our fuzzy model based on the feedback received in each trial.

Learning in the fuzzy model could happen in two ways: First, by adjusting the parameters of the membership functions, and second, by adjusting the consequents of the rules. Here, we only consider the later, but the method can be easily extended to adjusting the parameters of the membership functions.

In the neural network representation of Figure \ref{fig:Fuzzy_NN}, the consequents of the rules were modeled as the weights from the neurons of layer 2 to the upper neuron of layer 3. In the previous sections, we assumed that these weights are either 0 (if the consequent of the corresponding rule is ``continue") or 1 (if the consequent is ``stop"). Learning the consequent of the rules is equivalent to adjusting these weights. We assume that the subject has some heuristic if-then rules in mind at the beginning of the experiment. Based on these rules, the weights corresponding to the consequents are initialized to either 0 or 1 at the beginning of the experiment. The learning algorithm proposed next, will update these values after each trial and based on the received feedback. This is an appealing feature of the model: in contrast to many of the learning models which initialize the weights randomly, in our model the initial values of the weights represents the subject's prior and heuristic knowledge about how to perform the task.

To develop the learning algorithm, we first need to determine the objective of learning. Consider the experimental design explained in Section \ref{sec:experiment}. To make referencing easier, we explain this design here again. The experiment consisted of 40 blocks of trials with fixed duration. Since the blocks' duration was fixed the total number of trials a subject could experience during the experiment depended on her speed. Each trial could be either easy or hard. In the easy trials the pay-off is $\pm20$ coins while this pay-off is $\pm1$ coin in the hard trials. By responding faster, the subject could experience more trials and therefore increase her chance to achieve more coins. However, responding faster increases the chance of responding incorrectly which results in losing coins. A rational subject will try to maximize the number of coins she achieves during the whole trial. In this experimental design, this is equivalent to maximize the \textit{average reward rate}, the expected value of reward divided by the expected time to achieve that reward. To see this, suppose that a subject has balanced between her speed and accuracy such that her average reward rate is $\rho$. Since the total duration of the experiment is fixed, say $T$, the expected reward will be $T\cdot \rho$. The subject cannot control $T$, and so to maximize the total reward she has to maximize $\rho$. Therefore, we assume that the learning objective is to maximize the average reward rate.

In the experiment explained in Section \ref{sec:experiment}, a cue presented at the beginning of each trial would indicate the difficulty of the trial. Therefore, the subjects could use two different decision boundaries for the two conditions. Here, we develop the learning model for a slightly different version of this experiment. Specifically, we consider an experiment in which the cues are not presented. This was Experiment B in \cite{khodadadi_learning_2016}. In that paper, we showed that the optimal strategy is to use a single time-decreasing decision boundary for both easy and hard trials. In the learning model below we also assume that the subject tries to learn a single boundary for both types of trials. The extension to the case where the cues are presented and the subject uses two decision boundaries is straightforward. We first explain a general reinforcement learning (RL) algorithm and then discuss how it can be used to develop a learning algorithm for our problem. 

In the RL framework the problem is modeled as a Markov decision process (MDP). In an MDP a learning \textit{agent} interacts with its \textit{environment}. The environment is modeled as a set of \textit{states}. In each state the agent can choose an action form a set of possible actions in that state. After the action is chosen, the environment transitions to a new state and some reward (positive or negative) are delivered to the agent. Usually, both state transition and rewards are stochastic. The probability distribution of the state transitions and rewards for each action determine the \textit{dynamic of the environment}. In the RL framework, it is assumed that this dynamic is unknown to the agent. The agent's goal is to learn what action to choose in each state to maximize its total outcome. The mapping from states to actions (what action to choose in each state) is called the agent's \textit{policy}. Therefore, the goal is to learn the \textit{optimal policy}, the mapping that leads to the maximum total outcome. Both the action and state space could be either discrete or continuous. Also, the transition between the states could happen in discrete time steps or continuously. As we will show, our problem can be modeled as a continuous space, discrete action and time MDP, and so next we explain an RL algorithm for such MDPs.

Suppose that the agent is following a policy $\pi$. This policy specifies the probability of choosing each action in each state. It is desired to know ``how good" is this policy because then we can adjust it toward the optimal policy. One method to quantify the performance of a policy is through the \textit{action-state values}, $Q_\pi(s,a)$ defined as follows:

\begin{equation}
Q_\pi(s,a)=\textnormal{E}_\pi\bigg[\sum_{k=0}^{T-t} r_{t+k} \bigg| s_t=s,a_t=a\bigg]
\label{eq:q_value}
\end{equation}

We need to explain the notation used in this equation. First, we have assumed that the problem in hand is \textit{episodic}. In episodic tasks, the agent starts from an initial state, interacts with the environment for the maximum of $T$ time steps, and returns back to the initial state to start a new episode (\cite{sutton_reinforcement_1998}). In equation \ref{eq:q_value} it is assumed that at time step $t$ the agent was in state $s$ and took action $a$. The quantity $Q_\pi(s,a)$ is the value of taking action $a$ in state $s$, assuming that agent will follow policy $\pi$ afterwards. This value is defined as the expected sum of the rewards, $r_t$, that the agent will receive from time step $t$ to the final time step $T$. As we mentioned before, both state transitions and rewards are stochastic, and so the expectation in the equation should be computed with respect to both. However, since the dynamic of the environment is unknown to the agent, it is not possible to compute this expectation. The RL algorithms address this problem by approximating this expectation. 

When the state space is discrete, $Q_\pi(s,a)$ can be represented as a table with one cell for each state-action pair. Then the RL algorithm will learn one value for each pair. This is not possible, however, when the state space is continuous. One way to tackle this issue is to represent the action values as a parametric function of the continuous state, $Q_{\pi,\theta}(s,a)$, where $\theta$ represents the parameters. Then, the RL algorithm will learn the parameters such that $Q_{\pi,\theta}(s,a)$ is a good approximation of the true values of $Q_\pi(s,a)$. Here, we take this approach. 

A popular RL algorithm for updating the parameters is based on the gradient-descent method and updates the parameters after each time step $t$ as follows (see \cite{sutton_reinforcement_1998}, Chapter 8 for more details):

\begin{equation}
\theta_{t+1}=\theta_{t} - \alpha \cdot \big( v_t -  Q_{\pi,\theta_t}(s_t,a_t) \big) \cdot \nabla_{\theta_t} Q_{\pi,\theta_t}(s_t,a)
\label{eq:update_theta}
\end{equation}

\noindent In this equation, $\alpha$ is the learning rate, and $ \nabla_{\theta_t} Q_{\pi,\theta_t}(s,a)$ is the gradient of the function $Q_{\pi,\theta_t}(s,a)$ with respect to the parameter vector $\theta_t$. Theoretically, $v_t$ should be the true action values $Q_\pi(s,a)$. However, as we explained before, the agent does not have access to these true values and therefore this is replaced by an approximation of the true values. For example in a variant of the RL algorithms called \textit{Q-learning}, $v_t$ is replaced by the one step prediction of the action value, $r_t+\max_a(Q_{\pi,\theta_t}(s_{t+1},a))$.

Now, we explain how this learning algorithm can be used to adjust the parameters of the proposed fuzzy model. To this end, we first need to specify the state and action spaces in our problem. We choose the state space to be the two dimensional space of the elapsed time in a trial and the canoe position at each time. This is obviously a continuous state space. At the beginning of each trial, the environment is at state (0,0). After that the state at each moment is determined by the canoe position and the elapsed time until the subject responds. At this point, the reward is delivered and one episode of the task is finished and the environment returns to the initial state (0,0). At each state, the subject has two choices: to continue accumulating information or to stop and respond. Therefore, he action space a each state is the discrete set $\{continue,stop\}$.

Now, we make two simplifying assumptions that reduce the computational cost of the model greatly. First, we assume that the parameters of the fuzzy model are updated only at the beginning of each trial. Second, we assume that at state (0,0) the subject always chooses the action ``continue". Based on the first assumption, Equation \ref{eq:update_theta} is used only once in each trial to update the parameters (instead of using it at each moment $t$ within a trial). Also, we will have $s_t=s_{t+1}=(0,0)$ in this equation. Therefore, if we use the Q-learning updating rule we have:

\begin{equation}
v_t-Q_{\pi,\theta_t}(s_t,a_t)=r_t+\max_a(Q_{\pi,\theta_t}((0,0),a))-Q_{\pi,\theta_t}((0,0),a_t)
\label{eq:TDerror}
\end{equation}

Based on the second assumption, there is only one possible action at state (0,0), $a=``coninue"$, and so the last two terms in Equation \ref{eq:TDerror} will cancel out, and we will have $v_t-Q_{\pi,\theta_t}(s_t,a_t)=r_t$, where $r_t$ is the reward achieved in the current trial.

The resulting learning rule is based on the definition of the action values in Equation \ref{eq:q_value}. In this definition, we have only considered the rewards and so the resulting learning rule will try to maximize the expected reward in each trial. However, as we mentioned, this is not the objective in the experimental design we explained at the beginning of this section. Instead, the objective is to maximize the expected reward rate. \cite{das_solving_1999} showed that this objective can be achieved by replacing $r_t$ with $r_t-\rho_t \cdot d_t$, where $\rho_t$ is the current estimate of the reward rate, and $d_t$ is the time spent to achieve reward $r_t$. For our experiment, $r_t$ is the reward achieved in a trial and therefore $d_t$ is the total time spent on that trial. The term $\rho_t \cdot d_t$ can be considered as the cost of time: since the time is limited, every moment that is spent on a trial is equivalent to losing the opportunity to spent that time on other trials, which on average result in $\rho_t$ units of reward per time unit.

\begin{figure}[!h]
	\centering
	\includegraphics[width=1\linewidth]{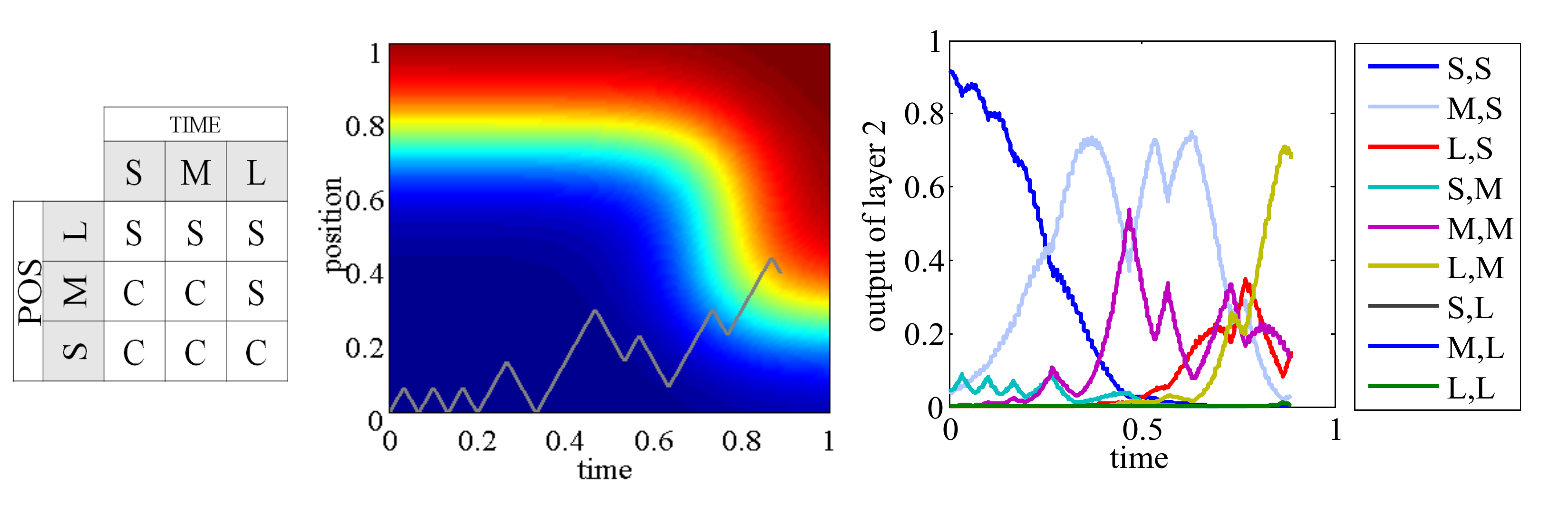}	
	\caption{{\bf Simulation of one trial of canoe task with fuzzy model.} The left panel is the rule-base of the fuzzy model used for this simulation The right panel shows the output of this model for a range of input variables. A simulated canoe path is also shown in this figure. The range of the input variables has been normalized. The subject has responded at time about 0.9 and when the canoe position was about 0.4. The right panel shows the output of the neurons of layer 2. These values show how much each rule was active during the trial. As it can be seen, different rules were dominant at different points in the trial. The legend of this figure shows the premise of each rule. For example `M,S' corresponds to the rule with the following premise: ``if time is M and position is M".} 
	\label{fig:simulatedTrial_RL}
\end{figure}

There is one more problem with the learning rule developed so far, namely the ``credit assignment problem". This can be best explained through an example. Consider a fuzzy model with the rule-base shown in the left panel of Figure \ref{fig:simulatedTrial_RL}. The output of this model for a range of the input variables is shown in the middle panel of this figure (both variables are normalized so their range is [0,1]). To compute this output we have set the center of the membership functions at 0, 0.5 and 1 for both input variables and set their variance at 0.04. A simulated canoe path is also shown in this figure. Suppose that the subject's response in this trial was correct and the value of $\rho$ is such that $r-\rho \cdot d>0$ (where $d$ is the reaction time which is about 0.9), and therefore, the subject has performed relatively well in this trial. The goal is to update the consequents of the rules, or equivalently, the weights from layer 2 to the upper neuron of layer 3 in the neural network implementation of the model (Figure \ref{fig:Fuzzy_NN}). The good performance of the subject in this trial is not due to the consequent of one rule, instead, all rules which have been activated in this trial should take the credit. This is the credit assignment problem: since, the reward is delivered at the end of each trial, and because we update the consequents only once and at the beginning of the next trial, the rules which have been activated in the middle of the trial will not receive the credit they deserve. For example, in the simulated trial shown in Figure \ref{fig:simulatedTrial_RL}, part of the credit should go to the consequent of the rule ``if time is S and position is S then continue". Because, if the consequent of this rule was ``stop" then the subject might have stopped accumulating information sooner. As we explained in Section \ref{sec:fit_fuzzy}, the output of the neurons of layer 2 encodes how much each rule has been activated for the given values of the input variables. These outputs for the simulated trial are shown in the right panel of Figure \ref{fig:simulatedTrial_RL}. As it can be seen, several rules have been activated at different points within the trial. At the time the subject responded, the rule ``if time is L and position is M then stop" was dominant. However, before time 0.6 this rule was not active at all and other rules were controlling the behavior. Therefore, these rule should get some of the credit for the performance in this trial. 

Here, we consider a simple solution to the credit assignment problem: we assume that the subject has a memory of the activation of each rule and at the time of updating, each rule is updated proportional to its activation during the trial. Let $y_i(t)$ denote the output of the $i^{th}$ neuron of layer 2 (which encodes how much rule $i$ is active) at time $t$ within a trial. Then we define a memory variable $z_i(t)$ as follows:

\begin{equation} 
z_i(t+\Delta t)= \gamma \cdot z_i(t) + y_i(t+\Delta t) \, \, , z_i(0)=0
\end{equation} 

\noindent where $\gamma$ is the memory leakage. 

On the other hand, using the method explained in the appendix, we can show that $\frac{\partial Q_{\pi}(s_t,a)}{\partial \theta_j}=\frac{y_{(2,j)}}{y_{(3,2)}}$, where $y_{(2,j)}$ is the output of the neuron $j$ of layer 2 and $y_{(3,2)}$ is the output of the second neuron of layer 3. But, since we have normalized the membership functions we have $y_{(3,2)}=1$ and so $\frac{\partial Q_{\pi}(s_t,a)}{\partial \theta_j}=y_{(2,j)}$. Therefore, we can replace $\nabla_{\theta_t} Q_{\pi,\theta_t}(s_t,a)$ with the vector of the activations of the neurons of layer 2, or with the memory of them. The final form of the reinforcement learning rule is as follows:

\begin{equation}
	\theta_j(k+1)=\theta_j(k) - \alpha \cdot \big( r_k - \rho \cdot d_k \big) \cdot z_j(k)
	\label{eq:update_theta_final}
\end{equation}

\noindent In this equation, $k$ is the trial number and $z_j(k)$ is the value of the memory for neuron $j$ in layer 2 at the end of trial $k$.

\begin{figure}[!h]
	\centering
	\includegraphics[width=1\linewidth]{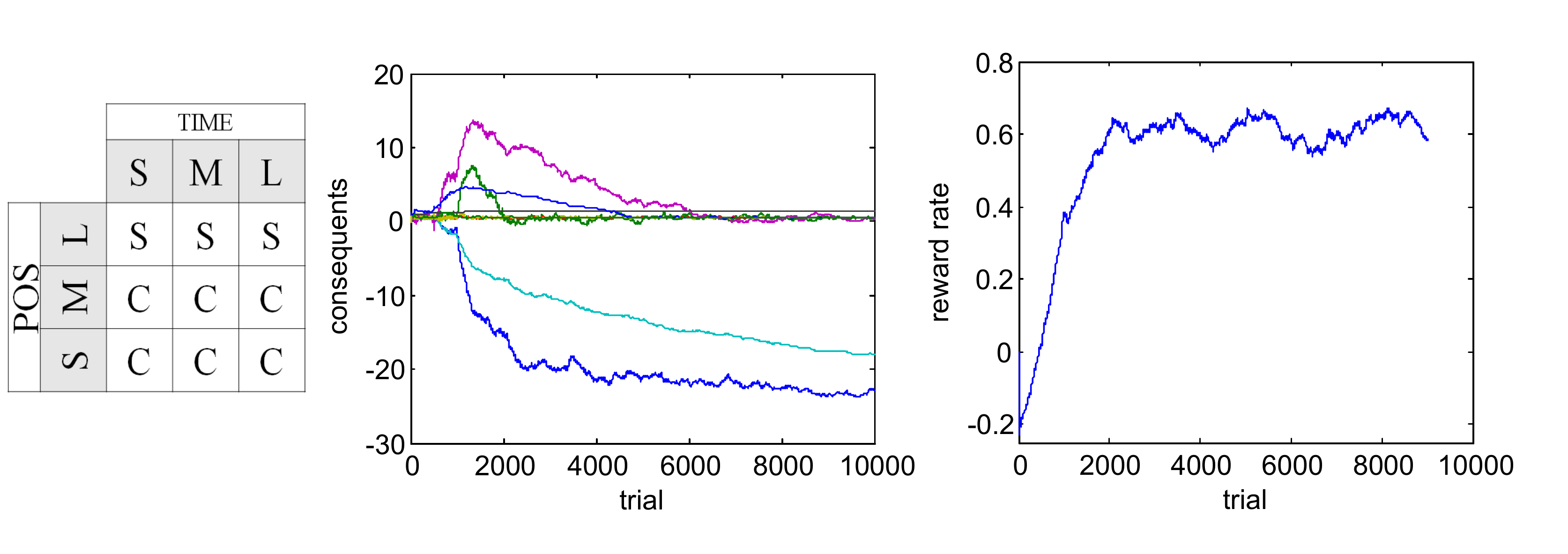}	
	\caption{{\bf Results of simulation of the fuzzy model with RL algorithm.} Left: the rule-base of the model before learning. Middle: values of the consequent of the rules as a function of the trial number. Right: the average reward rate as a function of the trial number.} 
	\label{fig:RL_sim}
\end{figure}

Figure \ref{fig:RL_sim} shows the results of a simulation of this model. The left panel of this figure shows the rule-base of the fuzzy model at the beginning of the experiment and prior to learning. The middle panel of the figure shows how the parameters (the consequent of the rules) are adjusted throughout the simulation. The right panel shows the average reward rate as the function of the trial number. Since there is a lot of variability in the experiment, the estimate of the average reward was very noisy. Therefore, we reduced the learning rate $\alpha$ and ran the simulation for 1000 trials. Then, the average reward rate was estimated using a moving average method with the window size of 1000. As it can be in the figure, the average reward rate obtained by the initial model was about -0.2. However, the reinforcement learning algorithm learns to adjust the consequent of rules such that the average reward becomes about 0.6 after about 3000 trials. This value is much less than the optimal reward rate for this problem. In \cite{khodadadi_learning_2016} we showed that the optimal value of the reward rate is about 0.9. Our goal here was not to develop an algorithm which is able to find the optimal solution. Instead, we tried to show how an RL algorithm can be developed for the fuzzy model. More research is necessary to develop RL algorithms which can reach the optimal performance. More importantly, these models should be compared to the human subjects' learning behavior.

\begin{appendices}
\numberwithin{equation}{section}

\begin{center}
	\section{}
\end{center}
	
In this appendix, we derive the formulas for estimating the parameters of the fuzzy model for each subject. To this end, we will use the back-propagation method for the neural network implementation of the fuzzy model (Figure \ref{fig:Fuzzy_NN}). This network has four layers. We denote the output of layers 1 to 4 by $y_{(1,i)}, y_{(2,j)}, y_{(3,i)}$ and $y_{(4)}$, respectively, where for example $y_{(1,i)}$ is the output of neuron $i$ of layer 1.

As it was explained in Section \ref{sec:param_esimate}, in order to estimate a free parameter $\theta_i$ in the network using gradient descent, we need to evaluate $\frac{\partial L}{\partial \theta_i}$. Our goal is to estimate the mean and the variance of the membership functions. For example consider the mean $\mu_i$. We have:

\begin{equation}
y_{(1,i)}=\exp\bigg(-\frac{(x-\mu_i)^2}{\sigma_i^2}\bigg)
\label{eq:A1}
\end{equation}
	
Using the chain rule we have:

\begin{equation}
\frac{\partial L}{\partial \mu_i}=\frac{\partial L}{\partial y_{(1,i)}}\cdot \frac{\partial y_{(1,i)}}{\partial \mu_i}=\delta_{(1,i)} \cdot \bigg[2 \cdot  \frac{x-\mu_i}{\sigma_i^2} \cdot y_{(1,i)} \bigg]
\label{eq:A2}
\end{equation}

Similarly for $\sigma_i$ we have:

\begin{equation}
\frac{\partial L}{\partial \sigma_i}=\delta_{(1,i)} \cdot \bigg[2 \cdot  \frac{(x-\mu_i)^2}{\sigma_i^3} \cdot y_{(1,i)} \bigg]
\label{eq:A3}
\end{equation}

Therefore, to compute $\frac{\partial L}{\partial \theta_i}$ we need to evaluate $\delta_{(1,i)}$. This quantity is called the \textit{local gradient} (\cite{haykin_neural_1998}) and can be computed using the back-propagation method. In this method, the local gradients of a layer is computed using the local gradients of the layers on its right hand side. To simplify the notations we write the equations for only one training sample and so we have $L=y\cdot \log(y_{(4)}) + (1-y) \cdot \log(1-y_{(4)})$. We start from the last layer, layer 4. This layer has only one neuron and we have:

\begin{equation}
\delta_{(4)} = \frac{\partial L}{\partial y_{(4)}}= \frac{y}{y_{(4)}} - \frac{1-y}{1-y_{(4)}}
\label{eq:A4}
\end{equation}

\noindent The next layer, layer 3, has 2 neurons and we have:

\begin{equation}
\delta_{(3,k)} = \delta_{(4)} \cdot \frac{\partial y_{(4)}}{\partial y_{(3,k)}}
\label{eq:A5}
\end{equation}

\noindent Also as it can be seen in Figure \ref{fig:Fuzzy_NN} we have:

\begin{equation}
y_{(4)}=\frac{y_{(3,1)}}{y_{(3,2)}}
\label{eq:A6}
\end{equation}

\noindent and therefore:

\begin{equation}
\frac{\partial y_{(4)}}{\partial y_{(3,1)}} = \frac{1}{y_{(3,2)}} \, \, , \, \, \frac{\partial y_{(4)}}{\partial y_{(3,2)}} = -\frac{y_{(3,1)}}{y_{(3,2)}^2}
\label{eq:A7}
\end{equation}	

Next, we compute te local gradients for layer 2. The number of neurons in this laer is equal to $r$ te number of rules in the rule-base. For each neuron of this layer we have:

\begin{equation}
\delta_{(2,j)} = \sum_k \delta_{(3,k)} \cdot \frac{\partial y_{(3,k)}}{\partial y_{(2,j)}}
\label{eq:A8}
\end{equation}

\noindent As it can be seen in Figure \ref{fig:Fuzzy_NN}, we have:

\begin{equation}
y_{(3,1)} = \sum_j c_j \cdot y_{(2,j)}  \, \, , \, \, y_{(3,2)} = \sum_j y_{(2,j)}
\label{eq:A9}
\end{equation}

\begin{equation}
\frac{\partial y_{(3,1)}}{\partial y_{(2,j)}} = c_j \, \, , \, \, \frac{\partial y_{(3,2)}}{\partial y_{(2,j)}} = 1
\label{eq:A10}
\end{equation}

Finally, we can compute $\delta_{(1,i)}$. We have:

\begin{equation}
\delta_{(1,i)} = \sum_j \delta_{(2,j)} \cdot \frac{\partial y_{(2,j)}}{\partial y_{(1,i)}}
\label{eq:A11}
\end{equation}

\noindent and so we need to compute $\frac{\partial y_{(2,j)}}{\partial y_{(1,i)}}$. For example consider $y_{(2,1)}$. We have:

\begin{equation}
y_{(2,1)} = y_{(1,1)} \cdot y_{(1,4)}
\label{eq:A12}
\end{equation}

\noindent and so:

\begin{equation}
\frac{\partial y_{(2,1)}}{\partial y_{(1,1)}}=y_{(1,4)}, \, \, \, \frac{\partial y_{(2,1)}}{\partial y_{(1,4)}}=y_{(1,1)}, \, \, \, \frac{\partial y_{(2,1)}}{\partial y_{(1,i)}}=0 \, \, \, for \, \, i\ne 1,4
\label{eq:A13}
\end{equation}

\noindent For other values of $i,j$ the term $\frac{\partial y_{(2,j)}}{\partial y_{(1,i)}}$ can be computed similarly.

The back-propagation algorithm is performed in two stages. For a given training data sample $(x,y)$, in the \textit{forward pass} we compute te output of each layer from left to right of the network, using Equations \ref{eq:A1}, \ref{eq:A12}, \ref{eq:A9} and \ref{eq:A6}. Then in the \textit{backward pass}, the local gradients are computed from the output layer to layer 1 using Equations \ref{eq:A5}, \ref{eq:A8} and \ref{eq:A11}. This procedure is performed several times until the estimated parameters converge.
	
\end{appendices}

\section*{References}
\bibliography{references}

\end{document}